\documentclass[%
reprint,
twocolumn, 
amsmath,amssymb,
aps,
]{revtex4}

\usepackage{color}
\usepackage{graphicx}
\usepackage{dcolumn}
\usepackage{bm}

\begin{document}

\preprint{APS/123-QED}

\title{Efficient Quantum Feature Extraction for CNN-based Learning}

\author{Tong Dou$^{1}$, Guofeng Zhang$^{2}$, and Wei Cui$^{1,}$}
\email{aucuiwei@scut.edu.cn}
\address{$^{1}$School of Automation Science and Engineering, South China University of Technology, Guangzhou 510640, China}
\address{$^{2}$Department of Applied Mathematics, The Hong Kong Polytechnic University, Hongkong, China}

\affiliation{}

%
%
%

\date{\today}

\begin{abstract}
Recent work has begun to explore the potential of parametrized quantum circuits (PQCs) as general function approximators.
In this work, we propose a quantum-classical deep network structure to enhance classical CNN model discriminability. The
convolutional layer uses linear filters to scan the input data. Moreover, we build PQC, which is a more potent function 
approximator, with more complex structures to capture the features within the receptive field. The feature maps are 
obtained by sliding the PQCs over the input in a similar way as CNN. We also give a training algorithm for the proposed model.
The hybrid models used in our design are validated by numerical simulation. We demonstrate the reasonable classification
performances on MNIST and we compare the performances with models in different settings. The results disclose that the model 
with ansatz in high expressibility achieves lower cost and higher accuracy.
\end{abstract}

\maketitle



\section{\label{sec:level1}Introduction}

Quantum Computers use the principles of quantum mechanics for computing, which are considered more powerful than 
classical computers in some computing problems. In view of the rapid progresses in quantum computing hardware \cite{arute2019quantum}, 
we are entering into so called NISQ era \cite{preskill2018quantum}. Noisy intermediate scale quantum (NISQ) devices will 
be the only quantum devices that can be reached in near-term, where we can only use a limited number of qubits with little 
error correction. Thus, developing a useful computational algorithm using the NISQ devices is the urgent task at present.

Quantum machine learning (QML) \cite{biamonte2017quantum} is gaining attention in the hope of speeding up machine 
learning tasks with quantum computers. Combing quantum computing and machine learning, QML attempts to utilize the power 
of quantum computers to achieve computational speedups or better performance for some machine learning tasks. Many quantum 
machine learning algorithms, such as qSVM \cite{rebentrost2014quantum}, qPCA \cite{lloyd2014quantum} and quantum Boltzmann 
machine \cite{amin2018quantum}, have been developed. 
Recently, several NISQ algorithms, such as quantum transfer learning \cite{mari2020transfer} and quantum kernel methods \cite{havlivcek2019supervised, schuld2019quantum}, 
have been proposed. On the other hand, parameterized quantum circuits \cite{benedetti2019parameterized} provide an 
another path for QML toward quantum advantages in NISQ era. Compared to conventional quantum algorithms, such as Shor's 
algorithm \cite{shor1999polynomial}, Grover's algorithm \cite{grover1997quantum}, HHL algorithm \cite{harrow2009quantum}, PQC 
based quantum-classical hybrid algorithms are inherently robust to noise and could benefit from quantum advantages, by taking 
advantage of both the high dimensional Hilbert space of a quantum system and the classical optimization scheme. Several 
popular related algorithms have been proposed, including variational quantum eigensolvers (VQE) \cite{peruzzo2014variational, wecker2015progress, mcclean2016theory} 
and quantum approximate optimization algorithm (QAOA) \cite{farhi2014quantum, hadfield2019quantum}.

In addition, quantum neural networks (QNNs) \cite{schuld2014quest, farhi2018classification, lloyd2020quantum} based on parameterized 
quantum circuits have been proposed. Some of QNN models utilized the thoughts of Convolutional Neural Network (CNN),
which is a popular machine learning model on the classical computer. Cong et al. \cite{cong2019quantum} designed a quantum 
circuit model with a similar hierarchy to classical CNNs, which dealt with quantum data and could be used to recognize 
phases of quantum states and to devise a quantum error correction scheme. The convolutional and pooling layers were 
approximated by quantum gates with adjustable parameters. Ref. \cite{henderson2020quanvolutional} proposed a new quanvolutional 
(short for quantum convolutional) filter, in which a random quantum circuit was deployed to transform input data locally. 
Quanvolutional filters were embedded into classical CNNs, forming a quantum-classical hybrid structure. In addition, a 
more complex QCNN design was presented in recent works \cite{Kerenidis2020Quantum, li2020quantum, zheng2021speeding}, where delicate 
quantum circuits were employed to accomplish quantum inner product computations and approximate nonlinear mappings of 
activation functions.

In this work, we propose a type of hybrid quantum-classical neural network based on PQCs for in analogy with the CNN. 
As a popular scheme in machine learning field, CNN replaces the fully connected layer with a convolutional layer. The 
convolutional layer only connects each neuron of the output to a small region of the input which is referred to as a feature 
map, thus greatly reducing  the number of parameters. 
However, the filter in classical CNN model is a generalized linear model (GLM). It is difficult for linear filters to 
extract the concepts are generally highly nonlinear function of the data patch. In Network-in-Network (NiN) \cite{lin2013network}, the linear 
filter replaced with a multilayer perceptron which is a general function approximator. As a ``micro network'', multilayer 
perceptron can improve the abstraction ability of the model. In the field of QML, PQCs are considered as the ``quantum network'' structures. 
Recent works \cite{schuld2021effect, goto2021universal,liu2021hybrid} shown that there exist PQCs which are universal function approximators 
with a proper data encoding strategy. Combining these ideas, we replace the linear filter with a PQC. The resulting structure 
is called the quantum feature extraction layer.

The key idea of our hybrid neural network is to implement the feature map in the convolutional layer with quantum parameterized circuits, and correspondingly, 
the output of this feature map is a correlational measurement on the output quantum state of the parameterized circuits. 
Different from Ref. \cite{henderson2020quanvolutional}, which uses parameters fixed circuits, we can iteratively update the 
parameters of circuits to get better performance. We also give the training algorithm of this hybrid quantum-classical 
neural network which are similar to backpropagation (BP) algorithm, meaning that our model is trained as a whole. Hence 
it can be trained efficiently with NISQ devices. Therefore, our scheme could utilize all the features of classical CNN, 
and is able to utilize the power of current NISQ processors.
Notice that the methods of data encoding and decoding are different from \cite{henderson2020quanvolutional}.

This paper is organized as follows. Section II is the preliminary, in which we first review the classical convolution 
neural network and the framework of PQCs. Then, the proposed quantum-classical hybrid model 
is described in detail in Section III. In Section IV, we present numerical simulation where we apply the hybrid model to 
image recognition and analyze the result. Conclusion and discussion are given in Section V.


\section{Preliminaries}

\subsection{Supervised learning: construct a classifier}

In a supervised classification task, we are given a training set $T$ and a test set $S$ on a label set $C$. Assuming 
there exists a map $m:T \cup S \rightarrow C$ unknown to the algorithm. Our task is to construct a model to infer an 
approximate map on the test set $\tilde{m}:S \rightarrow C$ only receiving the labels of the training set such that the 
difference between $m(s)$ and $\tilde{m}(s)$ is small on the test set $s \in S$. For such a learning task to be meaningful 
it is assumed that there is a correlation between the labels given for training set and the true map. A popular approach 
to this problem is to use neural networks(NNs), where the training data goes through the linear layer and the activation 
function repeatedly to map the true label. After constructing a NNs model that is accurate enough, we can perform the 
prediction of test set. 

\subsection{Convolution neural network(CNN)}
CNN dates back decades \cite{lecun1989backpropagation, lecun1998gradient} which is a specialized kind of neural network 
for processing data that has a known grid-like topology. Deep CNNs have recently shown an explosive popularity partially 
due to its success in image classification \cite{krizhevsky2012imagenet}. It also has been successfully applied to other 
fields, such as objection detection \cite{redmon2016you}, semantic segmentation \cite{long2015fully} and pedestrian detection \cite{zhang2016faster}, 
even for natural language processing(NLP) \cite{vaswani2017attention}.

Though there are several architectures of CNN models such as LeNet \cite{lecun1998gradient}, AlexNet \cite{krizhevsky2012imagenet},
GoogLeNet \cite{szegedy2015going}, VGG \cite{simonyan2014very}, ResNet \cite{he2016deep} and DenseNet \cite{huang2017densely}, they
use  three main types of layers to build CNN models: convolutional layer, pooling layer and fully-connected layer. Broadly, 
a CNN model can be divided into two parts: feature extraction module and classifier module. The former includes convolutional 
layer and pooling layer while the classifier module consists of fully-connected layers.

Formally, a convolutional layer is expressed as an operation $f_1$:
\begin{eqnarray}
f_1 (\textbf{X}) = g(\textbf{W}_1 * \textbf{X}+\textbf{B}_1),
\end{eqnarray}
where $\textbf{W}_1$ and $\textbf{B}_1$ represent the filters weights and biases respectively, and `$*$' denotes the 
convolution operation. Concretely, $\textbf{W}_1$ corresponds to $n_c$ filters of support $f\times f \times c$, where $c$ 
is the number of channels in the input $\textbf{X}$, $f$ is the spatial size of a filter. Intuitively, $\textbf{W}_1$ 
applies $n_c$ convolutions on the input, and each convolution has a kernel size $f\times f \times c$. The output is 
composed of $n_c$ feature maps. $\textbf{B}_1$ is a $n_c$-dimensional vector, whose each element is associated with a 
filter.  $g(\cdot)$ is an activation function which element-wisely apply a nonlinear transformation ,such as ReLU or Tanh, 
on the filter responses. 

A pooling layer is generally added after the convolutional layer. Its function is to progressively reduce the spatial size 
of the representation to reduce the amount of parameters and computation in the network. Two commonly used pooling operations 
are max pooling and average pooling. The former calculates the maximum value for each patch of the feature map while the 
latter calculates the average value for each patch on the feature map.  And the units in a fully-connected layer have full 
connections to all activations in the previous layer, as seen in regular neural networks.The activations are computed with 
a matrix multiplication followed by a bias offset, as shown in the operation $f_2$:
\begin{eqnarray}
f_2 (\textbf{x}) = h(\textbf{W}_2  \textbf{x}+\textbf{B}_2),
\end{eqnarray}
where  $\textbf{x}$,  $\textbf{W}_2$, $\textbf{B}_2$ are a input vector,  a weight matrix and a bias vector, respectively. 
$h(\cdot)$ is an activation function like $g(\cdot)$ mentioned.

\subsection{General framework of PQCs}

Parametrized quantum circuits, also known as ``variational circuits", are a kind of quantum circuits that have trainable 
parameters subject to iterative optimizations. Previous results show that such circuits are robust against quantum noise 
inherently and therefore suitable for the NISQ devices. Recently, there have been many developments in PQCs-based algorithms 
which combine both quantum and classical computers to accomplish a particular task. In addition to VQE and QAOA, the PQCs 
framework has been extended for applications in generative modeling \cite{dallaire2018quantum}, quantum data compression \cite{romero2017quantum}, 
quantum circuit compiling \cite{sharma2020noise} and so on.

Algorithms involving PQCs usually work in the following way. First, prepare a initial state $|\varphi_0\rangle$ by encoding 
input into the quantum device. Second, we need to choose an appropriate ansatz, that is, designing the circuit structure 
of a PQC $U(\bm{\theta})$, and apply $U(\bm{\theta})$ to $|\varphi_0\rangle$, where $\bm{\theta}$ is parameters of the 
circuit. Then measure the circuit repeatedly on a specific observable $\hat{O}$ to estimate an expectation value $\langle\hat{O}(\bm{\theta})\rangle$. 
Based on the $\langle\hat{O}(\bm{\theta})\rangle$ which is fed into a classical optimizer, we compute a cost function $L( \langle\hat{O}(\bm{\theta})\rangle)$ 
to be minimized by updating $\bm{\theta}$.

Several factors are of central importance in the PQC algorithms. First, it is necessary to find a suitable $\hat{O}$ for 
the given problem. That is, we require the minimum of $L( \langle\hat{O}(\bm{\theta})\rangle)$ to correspond to the solution 
of the task. How easily $\hat{O}$ can be measured on a quantum computer and the locality of $\hat{O}$(i.e. the number of 
qubits it acts non-trivially on) are also relevant.

Another aspect that determines the success of a PQC algorithm is the choice in ansatz for $U(\bm{\theta})$. While discrete 
parameterizations are possible, $\bm{\theta}$ usually are continuous parameters, such as gate rotation angles, in a PQC. 
In general, a PQC is of form
\begin{eqnarray}
U(\bm{\theta})=\prod_{j=1}^{N}U_{j}(\theta_{j})W_{j},
\end{eqnarray}
where $\bm{\theta}=(\theta_{1}, \dots, \theta_{N})$ are parameters, and $\{W_{j}\}_{j=1}^{N}$ is a set of unitaries without 
parameters while $U_{j}=e^{-i\theta_{j}V_{j}}$ is a rotation gate of angle $\theta_{j}$ generated by a Hermitian operator 
$V_{j}$ such that $V_{j}^{2}=I$. The rotation angles $\{\theta_{j}\}$ are typically considered to be independent.

\begin{figure*}[htbp]
    \begin{center}
    \footnotesize
    \begin{tabular}{cccc}
    \includegraphics[scale=0.3]{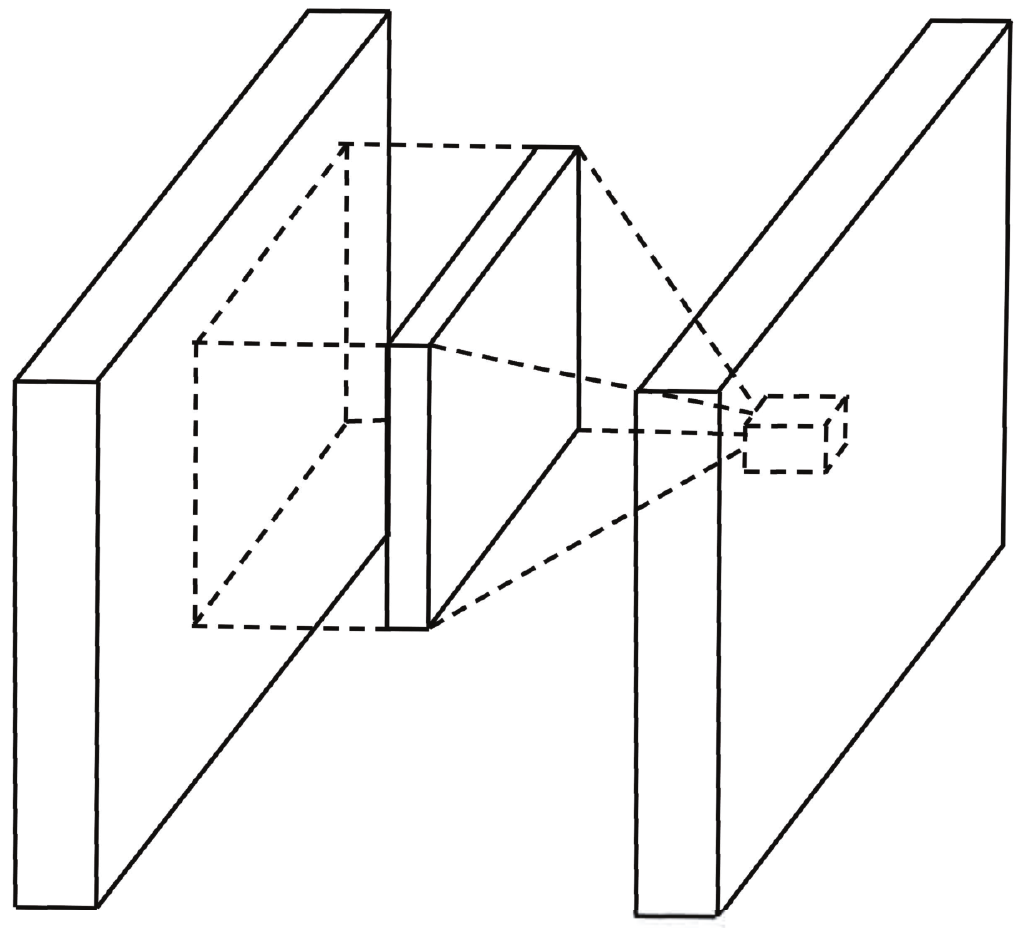} &
    \includegraphics[scale=0.375]{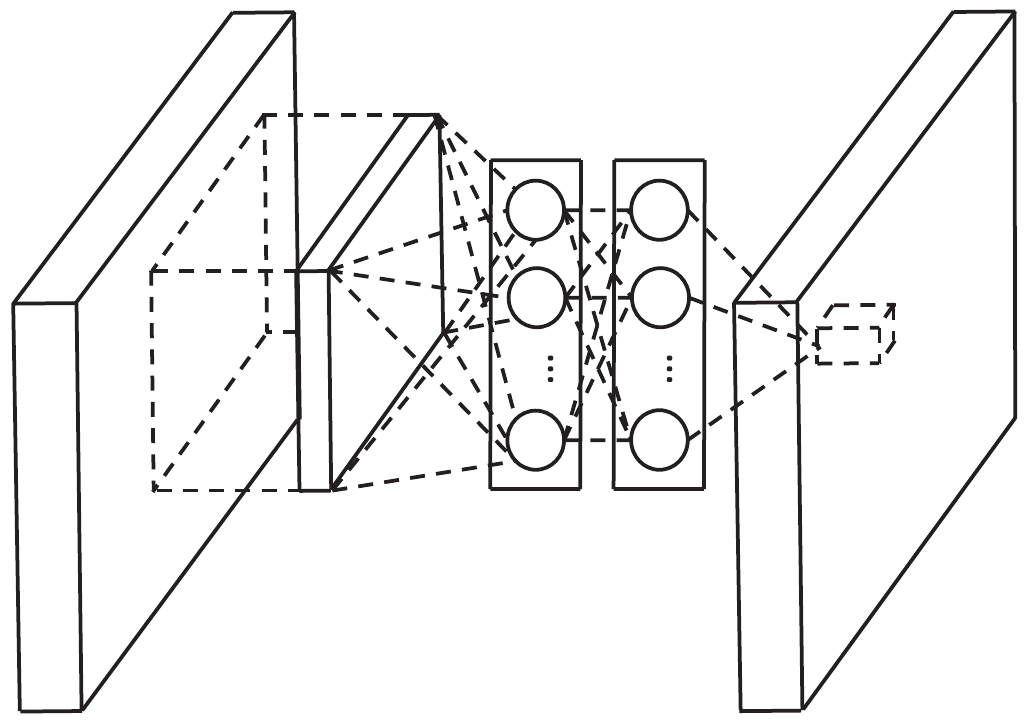} &
    \includegraphics[scale=0.375]{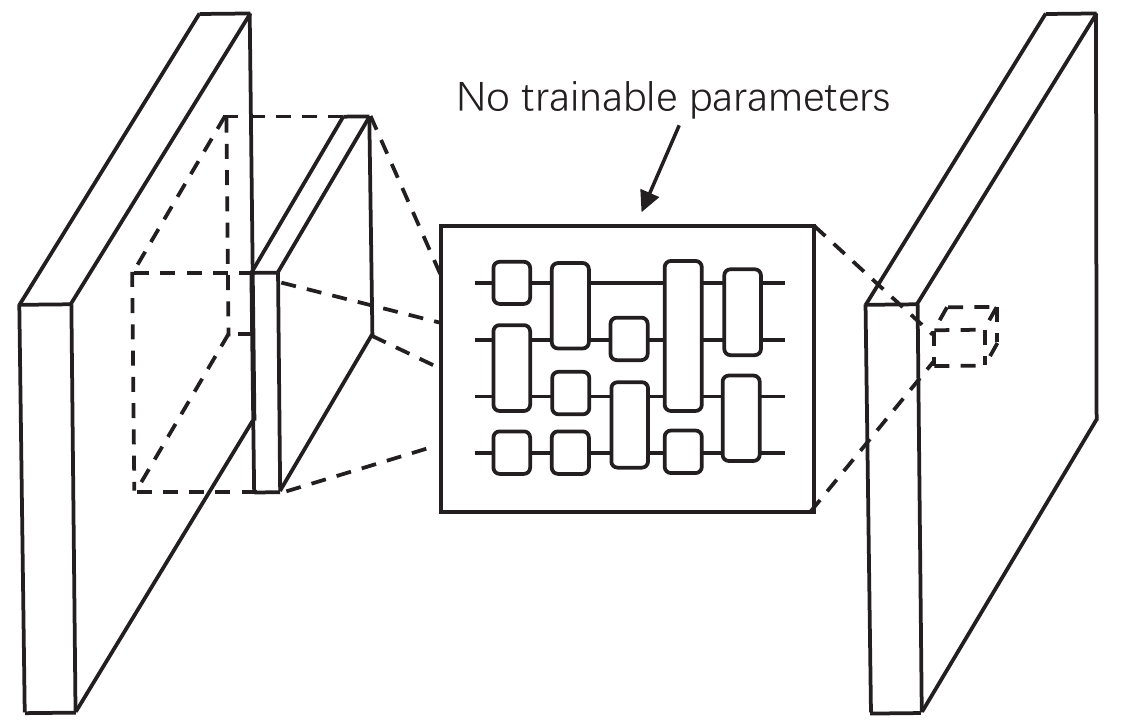} &
    \includegraphics[scale=0.375]{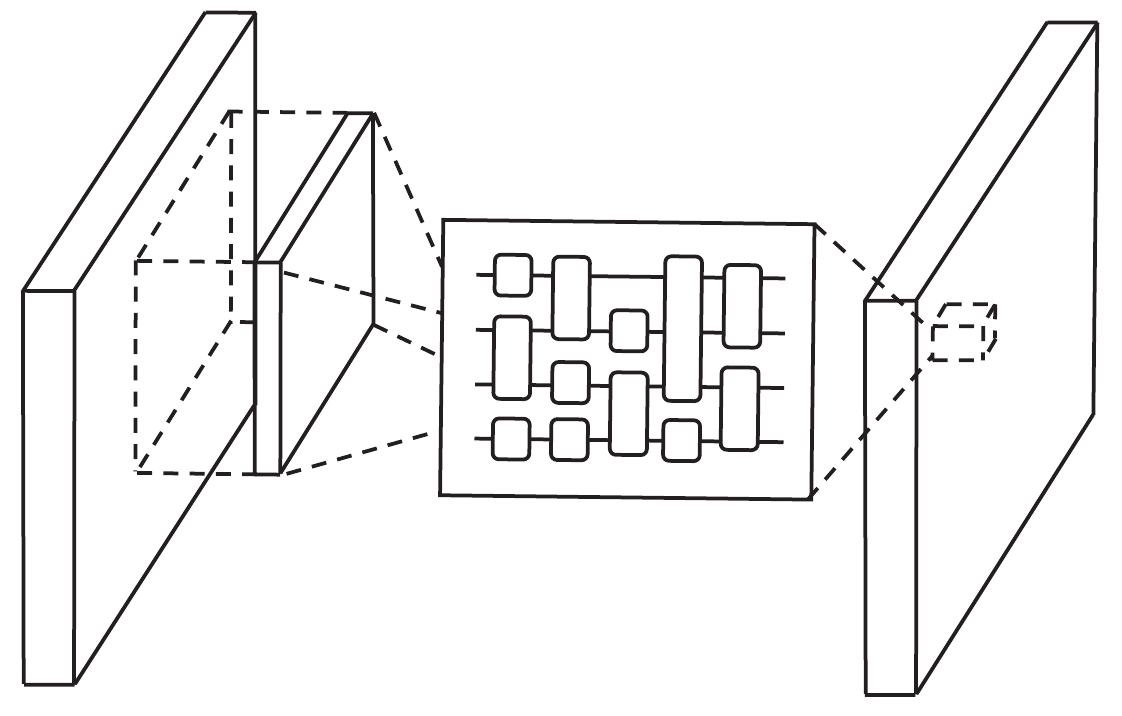} \\
    (a) convolutional layer & (b) Mlpconv layer & (c) Quanvolutional layer & (d) QFE layer \\
    \end{tabular}
    \end{center}
    \caption{Comparison of convolutional layer, mlpconv layer, quanvolutional layer and QFE layer. The 
    convolutional layer includes a linear filter. The mlpconv layer includes a multilayer perceptron. The quanvolutional layer 
    includes a random quantum circuit without trainable parameters. The QFE layer includes a ``micro quantum network''(we choose 
    PQC in this paper).}
    \label{fig:wide}
\end{figure*}

\section{Convolutional Network with Quantum Feature Extraction Layers}

In this section, we will first describe the quantum-classical hybrid model with quantum feature extraction layers and give 
a training algorithm. Then the possible experimental setup for its realization is discussed.

\subsection{Quantum Feature Extraction Layers} 
A quantum feature extraction(QFE) layer is a transformational layer using learnable parameteried quantum circuits. It is 
similar to a convolutional layer in classical CNNs. Combining pooling layers, fully-connected layer even with convolutional 
layer, we can construct a hybrid quantum-classical neural network. 

Suppose a QFE layer learns a mapping $F$, which conceptually consists of three parts, namely encoding, variational evolution, 
decoding.
\begin{enumerate}
\item \emph{\textbf{Encoding.}} This operation encodes patches from the input data $\bm{x}$(or the output of previous layer) into 
a quantum circuit by applying $U_{en}(\bm{x})$ to the initialized states $|0\rangle^{n}$. Hereafter, we abbreviate $|0\rangle^{n}$ 
as $|0\rangle$ if there is no confusion. Note that the output state $|\bm{x}\rangle$ are characterized by input $x$.
\item \emph{\textbf{Variational evolution.}} In this part, we apply a parameterized unitary circuit $U_{var}(\bm{\theta})$ to the 
quantum state $|x\rangle$. $U_{var}(\bm{\theta})$ maps the quantum state $|\bm{x}\rangle$ onto another quantum state $|\bm{x};\bm{\theta}\rangle$ 
using a serial of quantum gates. This operation can introduce quantum properties such as superposition, entanglement and 
quantum parallelism.
\item \emph{\textbf{Decoding.}} This operation include two parts: measurement and classical post-processing. Firstly, at the end 
of the circuit, measurements are performed on an observable $\hat{H}$ to get an expectation value. Then after the classical 
post-processing, we get a output of one step in QFE layer. Analogously to a classical convolution layer, each output is mapped 
to a different channel of a single pixel value.
\end{enumerate}

The structure of a QFE layer is compared with convolutional , quanvolutional and mlpconv layer in Fig.~\ref{fig:wide}. 
Next we detail our definition of each operation.

\subsubsection{Encoding}
There are two popular strategy to encode classical data into quantum circuit, including the basis encoding and the amplitude 
encoding. The basis encoding method treats two basic states of a qubit, $|0\rangle$ and $|1\rangle$, as binary values of 
a classical bit, 0 and 1. While the amplitude encoding method uses the probability amplitudes of a quantum state to store 
classical data. As described below, to update the parameters effectively, we need to calculate the partial derivative of 
input. However, the two methods mentioned above are difficult to achieve it. To satisfy the condition, we treat a single 
input value as the angle of a parametrized rotation gate. Combining with parameter-shift rule\cite{mitarai2018quantum}, 
we can calculate the analytical gradients of input on quantum circuits. Formally, the encoding part is expressed as an 
operation $F_1$:
\begin{eqnarray}
F_{1}(\bm{x}) = U_{en}(\bm{x})|0\rangle=|\bm{x}\rangle.
\end{eqnarray}
Here $\bm{x}=(x_{1},x_{2},\dots,x_{n})^{\bm{T}}$ is the input patch of QFE layer where $n$ is the dimension of $\bm{x}$(usually 
in a $3\times3$ or $5\times5$ size) and represents the size of the quantum circuit. $U_{en}(\bm{x})$ is a unitary to embed 
$\bm{x}$ into a quantum circuit. For example, as we used in the experiments, $U_{en}(\bm{x})$ can expressed as:
\begin{eqnarray}
U_{en}(\bm{x})=\bigotimes_{i=1}^{n}R_{y}(x_{i}).
\end{eqnarray}

\subsubsection{Variational evolution}
The first part encodes the each of input patch $\bm{x}$ as a quantum state $|\bm{x}\rangle$. In the second part, applying 
a unitary $U_{var}(\bm{\theta})$, we map the state $|\bm{x}\rangle$ into another state $|\bm{x};\bm{\theta}\rangle$. 
Utilizing the power of quantum mechanics, such as superposition and entanglement, we hope that it can improve the usefulness 
of features. The operation of the second part is:
\begin{eqnarray}
F_{2}(\bm{x})=U_{var}(\bm{\theta})|\bm{x}\rangle=|\bm{x};\bm{\theta}\rangle,
\end{eqnarray}
where $U(\bm{\theta})$ is an ansatz characterized by the parameters $\theta$ and $\theta$ is a vector whose dimension 
depends on the type of ansatz. It is important to find an expressive ansatz that can achieve a big range of quantum states. 
More generally, it is possible to repeat $U_{var}(\bm{\theta})$ to reach more quantum states. But this can increase the 
complexity of the model, and thus demands more training time.

\begin{figure}
\includegraphics[scale=0.5]{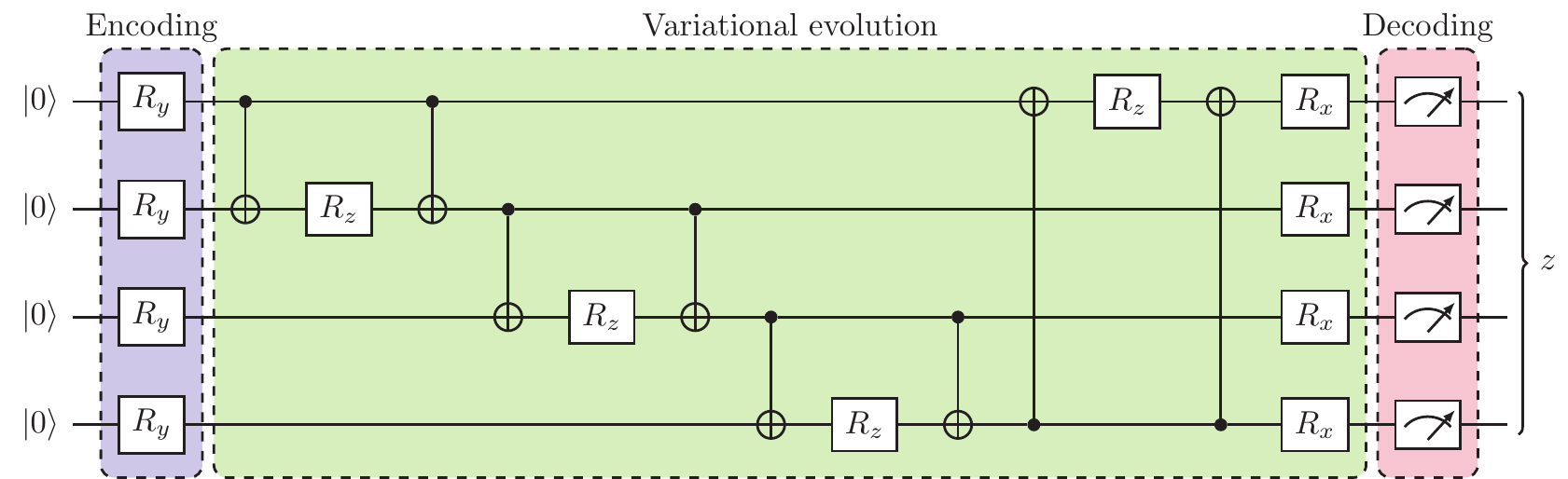}
\caption{\label{fig:epsart} The QAOA-heuristic circuit.}
\end{figure}

In order to implement the ansatz in a real quantum device,  we select $\{ H, R_{X}, R_{Y}, R_{Z}, CNOT\}$ to compose the ansatz
In this case, motivated by the universality of QAOA \cite{lloyd2018quantum, morales2020universality}, we use a QAOA-heuristic circuit as an ansatz. 
The circuit is as shown in Fig.~\ref{fig:epsart}.

\subsubsection{Decoding}
To access the information of the quantum states, we need to ``decode'' the quantum state $|\bm{x};\bm{\theta}\rangle$ into 
classical data, which consists of measurement and classical post-processing. At the measurement part, we measure a expectation 
value of a observable $\hat{H}$. Then, in classical post-processing part, we transform this expectation value to get the 
output by adding a bias $b$. The output is expressed as:
\begin{eqnarray}
F_3(\bm{x}) = \sigma(Tr(|\bm{x};\bm{\theta}\rangle\langle\bm{x};\bm{\theta}|\hat{H})+b),
\end{eqnarray}
where $|\bm{x};\bm{\theta}\rangle$ is the output state from the variational evolution part and $b$ is the bias. $\hat{H}$ 
is an observable composed of $\{ \hat{H_{k}} \}_{k=1}^{l} $ where $H_{k}$ is a fixed Hamiltonian. $Tr(\cdot)$ is the trace 
operator. $\sigma(\cdot)$ is the activation function. Its function is to manipulate the outputs into a range of $[0, 2\pi]$

\subsection{Backward pass of QFE layer}
In classical NNs, gradient descent based optimization algorithms are usually applied to update the parameters. A powerful 
routine in training  is the backpropagation algorithm, which combines the chain rule for derivatives and stored values 
from intermediate layers to efficiently compute gradients. However, this strategy can not generalized straightforwardly 
to the QFE layer, because intermediate values from a quantum circuit can only be accessed via measurements, which will cause 
the collapse of quantum states.
Using the parameter-shift rule and the chain rule, we proposed a training algorithm to compute the gradients of QFE Layer which 
can be embedded in the standard backpropagation. Hence we can train the hybrid model as a whole.

Suppose a QFE layer with parameters $\bm{\theta}$, $b$, whose input is $Z$ and output is $A$. The size of $Z$ is $m\times m$. 
The size of kernel is $f\times f$ and the stride is $s$, therefore, the size of $A$ is $(\lfloor \frac{m-f}{s} \rfloor +1) \times (\lfloor \frac{m-f}{s} \rfloor +1)$. 
Let $n=\lfloor \frac{m-f}{s} \rfloor +1$ to make the procedure clearer. Then, $Z$ and $A$ can be expressed as:
\begin{eqnarray}
Z = 
\begin{pmatrix}
z_{11} & z_{12} & \cdots & z_{1m} \\
z_{21} & z_{22} & \cdots & z_{2m} \\
\vdots & \vdots & \ddots & \vdots \\
z_{m1} & z_{m2} & \cdots & z_{mm}
\end{pmatrix},
\end{eqnarray}
\begin{eqnarray}
A = 
\begin{pmatrix}
a_{11} & a_{12} & \cdots & a_{1n} \\
a_{21} & a_{22} & \cdots & a_{2n} \\
\vdots & \vdots & \ddots & \vdots \\
a_{n1} & a_{n2} & \cdots & a_{nn}
\end{pmatrix}.
\end{eqnarray}

\begin{figure*}[htbp]
    \begin{center}
    \footnotesize
    \begin{tabular}{c}
    \includegraphics[scale=0.55]{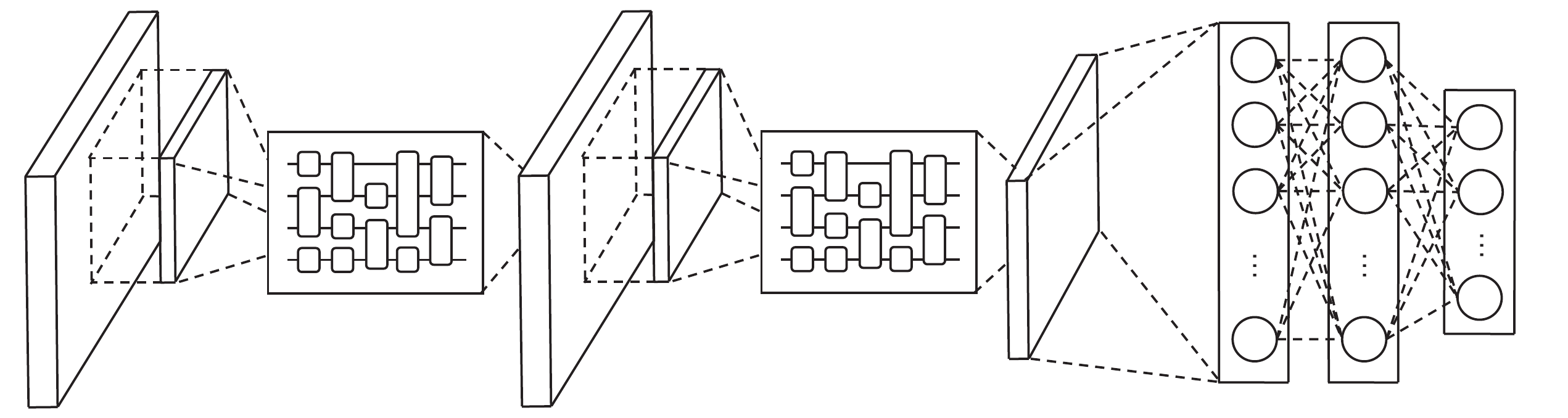} \\
    (a) The model that includes the stacking of two QFE layers and a classical ``classifier''. \\
    \\
    \includegraphics[scale=0.55]{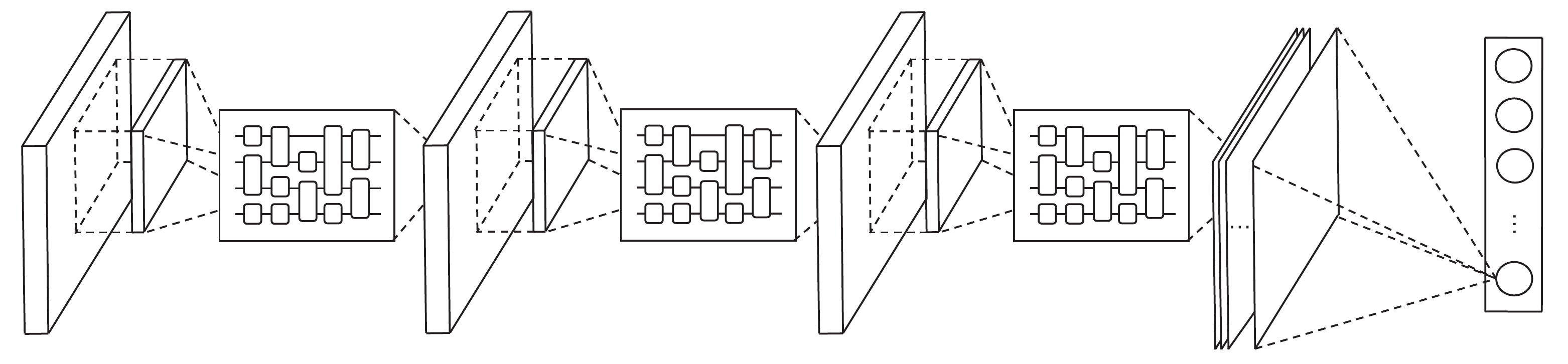} \\
    (b) The model that includes the stacking of three QFE layers and a global average layer. \\
    \end{tabular}
    \end{center}
    \caption{The overall structures of evaluated models.}
    \label{fig:3}
\end{figure*}

$L$ denotes the loss function. Firstly, consider the condition $s=1$ which is be commonly used. Assuming that the partial 
derivative 
$\frac{\partial L}{\partial A}$
was known. $\frac{\partial a_{ij}}{\partial z_{uv}}$ and $\frac{\partial a_{ij}}{\partial \theta_{k}}$ where $i,j=1,\cdots,n$, $u,v=1,\cdots,m$ 
and $k=1,\cdots,l$ can be calculated by parameter-shift rule.
To implement the backpropagation of the QFE layer, we need to get $\frac{\partial L}{\partial \bm{\theta}}$, $\frac{\partial L}{\partial b}$ 
and $\frac{\partial L}{\partial Z}$. For convenience, we flatten $\frac{\partial L}{\partial Z}$ and $\frac{\partial L}{\partial A}$ 
row by row as follow:

\begin{eqnarray}
\frac{\partial L}{\partial A}=
\begin{pmatrix}
\frac{\partial L}{\partial a_{11}} &  \cdots & \frac{\partial L}{\partial a_{1n}} \\
\frac{\partial L}{\partial a_{21}} &  \cdots & \frac{\partial L}{\partial a_{2n}} \\
\vdots & \ddots & \vdots \\
\frac{\partial L}{\partial a_{n1}} & \cdots & \frac{\partial L}{\partial a_{nn}} \\
\end{pmatrix}
\rightarrow
\frac{\partial L}{\partial \overline{A}}=
\begin{pmatrix}
\frac{\partial L}{\partial \overline{a}_{1}} \\
\frac{\partial L}{\partial \overline{a}_{2}} \\
\vdots \\
\frac{\partial L}{\partial \overline{a}_{n\times n}}
\end{pmatrix}.
\end{eqnarray}
For $\frac{\partial L}{\partial b}$, by the chain rule, we have
\begin{eqnarray}
\frac{\partial L}{\partial b}=\sum_{i,j}\frac{\partial a_{i,j}}{\partial b}\frac{\partial L}{\partial a_{i,j}}=\sum_{i,j}\frac{\partial L}{\partial a_{i,j}}=\sum\frac{\partial L}{\partial A}.
\end{eqnarray}
For $\frac{\partial L}{\partial \bm{\theta}}$, by the chain rule, we have
\begin{eqnarray}
\frac{\partial L}{\partial \theta_{k}} &=&\sum_{i=1}^{m}\sum_{j=1}^{m}\frac{\partial a_{ij}}{\partial \theta_{k}}\frac{\partial L}{\partial a_{ij}}=\sum_{i=1}^{m\times m}\frac{\partial \overline{a}_{i}}{\partial \theta_{k}}\frac{\partial L}{\partial \overline{a}_{i}} \\
& = & 
\begin{pmatrix}
\frac{\partial \overline{a}_{1}}{\partial \theta_{k}} & \frac{\partial \overline{a}_{2}}{\partial \theta_{k}} & \cdots & \frac{\partial \overline{a}_{m\times m}}{\partial \theta_{k}}
\end{pmatrix}
\begin{pmatrix}
\frac{\partial L}{\partial \overline{a}_{1}} \\
\frac{\partial L}{\partial \overline{a}_{2}} \\
\vdots \\ 
\frac{\partial L}{\partial  \overline{a}_{m\times m}}
\end{pmatrix}
=\frac{\partial \overline{A}}{\partial \theta_{k}}^{T}\frac{\partial L}{\partial \overline{A}}. \nonumber
\end{eqnarray}

Thus,
\begin{eqnarray}
\frac{\partial L}{\partial \bm{\theta}}=
\begin{pmatrix}
\frac{\partial L}{\partial \theta_{1}} \\
\frac{\partial L}{\partial \theta_{2}} \\
\vdots \\
\frac{\partial L}{\partial \theta_{l}} 
\end{pmatrix}
=
\frac{\partial \overline{A}}{\partial \bm{\theta}}^{T}\frac{\partial L}{\partial \overline{A}},
\end{eqnarray}
where 
$
\frac{\partial \overline{A}}{\partial \bm{\theta}}^{T}
=
\begin{pmatrix}
\frac{\partial \overline{a}_{1}}{\partial \theta_{1}} & \frac{\partial \overline{a}_{2}}{\partial \theta_{1}} & \cdots & \frac{\partial \overline{a}_{m\times m}}{\partial \theta_{1}} \\
\frac{\partial \overline{a}_{1}}{\partial \theta_{2}} & \frac{\partial \overline{a}_{2}}{\partial \theta_{1}} & \cdots & \frac{\partial \overline{a}_{m\times m}}{\partial \theta_{2}} \\
\vdots & \vdots & \ddots & \vdots \\
\frac{\partial \overline{a}_{1}}{\partial \theta_{l}} & \frac{\partial \overline{a}_{l}}{\partial \theta_{1}} & \cdots & \frac{\partial \overline{a}_{m\times m}}{\partial \theta_{l}} \\
\end{pmatrix}
$.

To compute the $\frac{\partial L}{\partial Z}$, we define $\frac{\partial A}{\partial Z}$ as
\begin{eqnarray}
\frac{\partial A}{\partial Z}=
\left (
\begin{array}{cccc}
\frac{\partial A}{\partial z_{11}} & \frac{\partial A}{\partial z_{12}} & \cdots & \frac{\partial A}{\partial z_{1m}}\\ 
\frac{\partial A}{\partial z_{21}} & \frac{\partial A}{\partial z_{22}} & \cdots & \frac{\partial A}{\partial z_{2m}}\\
\vdots & \vdots & \ddots & \vdots \\
\frac{\partial A}{\partial z_{m1}} & \frac{\partial A}{\partial z_{m2}} & \cdots & \frac{\partial A}{\partial z_{mm}}\\
\end{array}
\right ),
\end{eqnarray}
where
$
\frac{\partial A}{\partial z_{ij}}=
\begin{pmatrix}
\frac{\partial a_{11}}{\partial z_{ij}} & \frac{\partial a_{12}}{\partial z_{ij}} & \cdots & \frac{\partial a_{1n}}{\partial z_{ij}}\\
\frac{\partial a_{21}}{\partial z_{ij}} & \frac{\partial a_{22}}{\partial z_{ij}} & \cdots & \frac{\partial a_{2n}}{\partial z_{ij}}\\
\vdots & \vdots & \ddots & \vdots \\
\frac{\partial a_{n1}}{\partial z_{ij}} & \frac{\partial a_{n2}}{\partial z_{ij}} & \cdots & \frac{\partial a_{nn}}{\partial z_{ij}}\\
\end{pmatrix}
$. 

Then, the $\frac{\partial L}{\partial Z}$ can be expressed as
\begin{eqnarray}
\frac{\partial L}{\partial Z}=\frac{\partial A}{\partial Z} * \frac{\partial L}{\partial A},
\end{eqnarray}
where `$*$' denotes the convolution operation with the stride  $n$.

\section{Experiments}
In this section, we measure the performance of our hybrid models on MNIST \cite{lecun1998gradient} dataset by numerical 
simulations. All of the experiments are performed with Yao \cite{luo2020yao} package in Julia language.

\subsection{Dataset}
The MNIST dataset is composed of 10 classes of hand written digits 0-9. There are 60,000 training images and 10,000 testing images. 
Each image is a gray image of size $28 \times 28$. To save the training time, we cut the images into $22 \times 22$ without 
information loss, as the original images are $20 \times 20$. For training and testing procedure, we use 6000 and 600 samples 
of all categories evenly for training set and test set, respectively. The labels use one-hot encoding. 

\subsection{Effectiveness of proposed method}
\textbf{Architecture selection:} Here, to verify the feasibility of the scheme, we evaluated models of following architectures. The models all consist 
of stacked QFE layers instead of linear convolutional layers.
\begin{enumerate}
\item \emph{QFE layers with fully-connected layers.} A model, which is similar to the classical CNN, replaces the linear 
convolutional layers with QFE layers. The structure is as followed: QFE1 - POOL1 - QFE2 - POOL2 - FC1 - FC2 - FC3. As shown 
in the Fig.~\ref{fig:3}(a) In this model, the fully-connected layers work as a ``classifier''. We regard this model as model 1.
\item \emph{QFE layers with global average pooling.} A model uses global average pooling (GAP) instead of fully-connected 
layers at the top of the network, with following structure: QFE1 - POOL1- QFE2 - QFE3 - GAP. As shown in Fig.~\ref{fig:3}(b). 
In this case, GAP introduced as a regularizer in Ref. \cite{lin2013network} is used to remove the ``classical factors'' of 
the network. We regard this model as model 2.
\end{enumerate}
Each QFE layer uses ``quantum filters'' of size $3 \times 3$. Max pooling is used in the pooling layers which downsamples 
the size of input by a factor of 2.

\textbf{Training strategy:} For fair comparison, we adopt the same dataset for training. To save training time, we also 
explore a step-by-step training procedure. First, we train the network from scratch with a relatively small dataset. Then 
when the training is saturated, we add more training data for fine-tuning and repeat this step. With this strategy, the training converges 
earlier and faster than training with the larger dataset from the beginning. For initialization, the weights of QFE layers 
are initialized by drawing randomly from a uniform distribution with a range of $[-\pi, \pi]$, while the weights of fully-connected 
layers are initialized by a Gaussian distribution with zero mean and standard deviation 0.001. The models are trained 
using Adam optimization algorithm with mini-batches of size 50. The initial learning rate is 0.01 and is adjusted manually 
during training.

\subsection{Investigation of different settings}
To investigate the property of the model, we design a set of controlling experiments with different settings of two variables
- the type of the ansatz and the number of layers of ansatz, based on the structure of model 2 (i.e. QFE1 - POOL1- QFE2 - QFE3 - GAP).
\subsubsection{Type of ansatz}
Recent work by Sim etc.\cite{sim2019expressibility} studies expressibility the entangling capability of quantum circuits, 
showing these descriptors are related to the performance of PQCs-based algorithms. Here, we compare the model performance
with different ansaetze. Specifically, we choose circuit 1, 2, 9, 14 and 15 in \cite{sim2019expressibility} besides the 
QAOA-heuristic circuit.
\subsubsection{Number of layers}
In general, the performance would improve if we increase the number of layers of ansatz at the cost of running time. Here,
we examine the model sensitivity to different number of layers. In previous experiments, we set the number of layers of 
ansatz $L=3$ for each "quantum" filter. To be consistent with previous experiments, we keep the kinds of ansaetze but change the number 
of layers from 1 to 5. All the other settings remain the same with model 2.

Totally, we examine 30 models for the same training data with 2560 samples for 9 epochs. The experimental setup is shown in the Table I.

\begin{table}[htb] 
    \begin{center} 
    \caption{Experimental setup to investigate different settings} 
    \label{table:1} 
    \begin{tabular}{|c|c|c|} 
    \hline   \textbf{Epoch} & \textbf{Learning rate} & \textbf{batch size} \\ 
    \hline   1 & 0.01 & 32  \\ 
    \hline   2 - 3 & 0.005 & 32  \\ 
    \hline   4 - 6 & 0.001 & 32  \\
    \hline   7 - 9 & 0.0005 & 16  \\ 
    \hline 
    \end{tabular} 
    \end{center} 
    \end{table}

\subsection{Results}

\begin{figure}
\centering
\includegraphics[width=9cm,height=6.5cm]{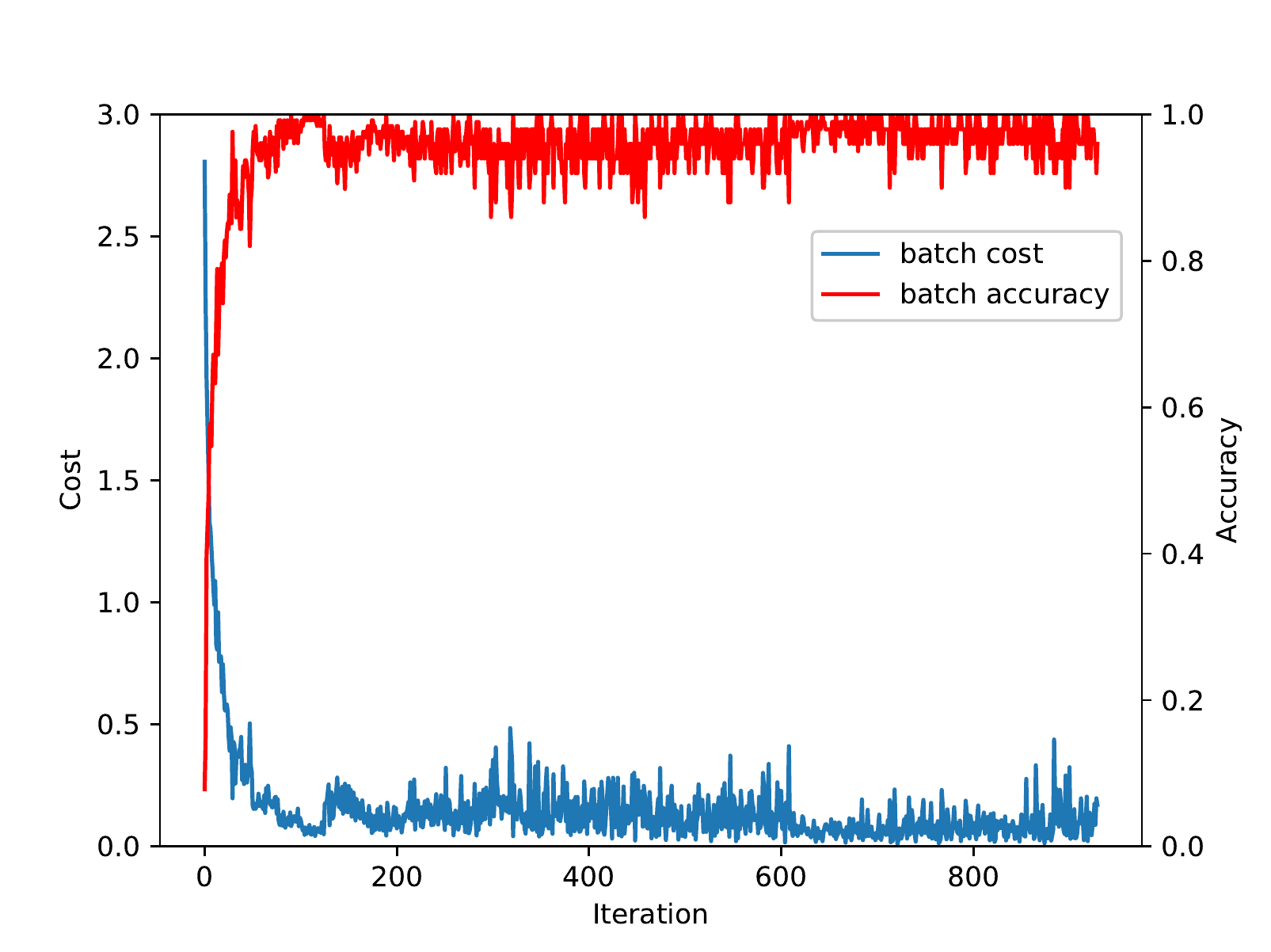}
\caption{\label{fig:model1} Training cost and accuracy for the model that includes QFE layers with fully-connected 
layers.}
\end{figure}
\begin{figure}
\centering
\includegraphics[width=9cm,height=6.5cm]{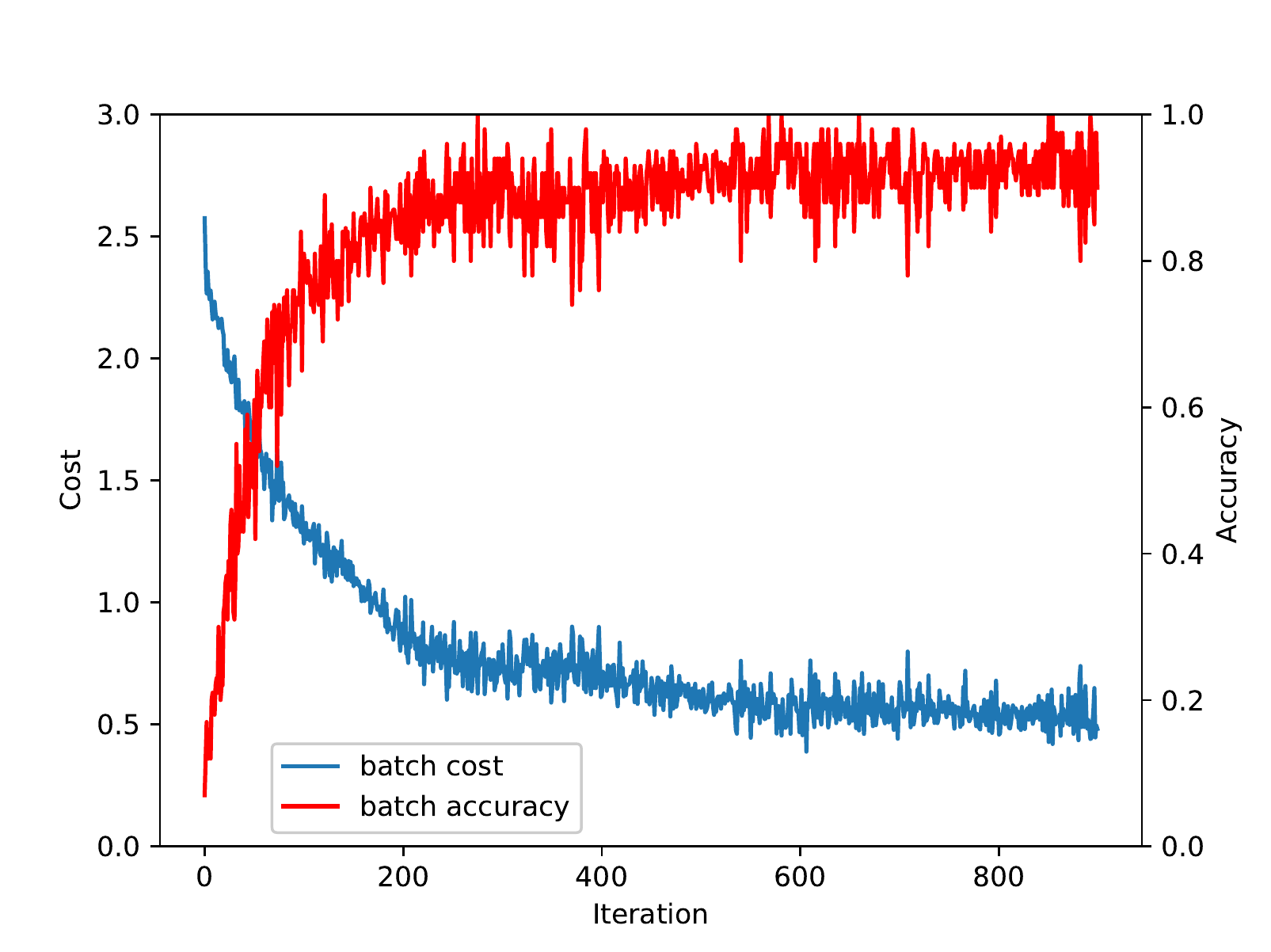}
\caption{\label{fig:model2} Training cost and accuracy for the model that includes QFE layers with GAP.}
\end{figure}

We first report results on classification to demonstrate that our method works with fully-connected layers or not. Then 
we compare the performance between different settings.

\textbf{Classification.} training performances between model 1 and model 2 are shown in Fig.~\ref{fig:model1} and Fig.~\ref{fig:model2}. For test set 
error rates, model1 and model2 achieve 3.1\% and 6.8\% respectively. It is reasonable for the toy models. However,
removing classical fully-connected layers, there is a small performance drop of 2 - 3\%. 

\textbf{Compare using different settings.} To further compare our approach with different settings mentioned above, we train the models 
with the same dataset and the results are shown in Fig.6. Note that these results are obtained using different ansaetze with 
different layers. We can observe that the model converges lower cost and achieves higher accuracy with deeper layers in ansatz. 
However, we note that ``saturation'' phenomenon exists, which means adding addition layers to PQC, the performance of the 
model does not always continue to improve. This phenomenon appears more and less in all of the ansaetze. Take circuit 9 as an example, 
the model with 2-layer PQC converges on a lower value and a higher accuracy than that with 1-layer PQC, while the curves of cost and accuracy 
are almost the same respectively, when the models equip with circuit 9 in 4 layers and 5 layers.
The ``saturation'' phenomenon may relate with expressibility saturation\cite{sim2019expressibility} or barren plateaus\cite{mcclean2018barren}.
To understand how expressibility correlates with the model performance further, future research is needed.

This section aims to demonstrate a proof of concept for the proposed method on a real-world dataset. 
Because the training is time-consuming, we construct our architectures based on the small model which has only two or three layers.
Note that model 2 which has no classical fully-connected layers achieves 6.8\% test set error. This shows the the 
effectiveness of QFC layers and their training algorithm.

\begin{figure*}[htbp]
  \begin{center}
  \footnotesize
  \begin{tabular}{ccc}
  \includegraphics[scale=0.27]{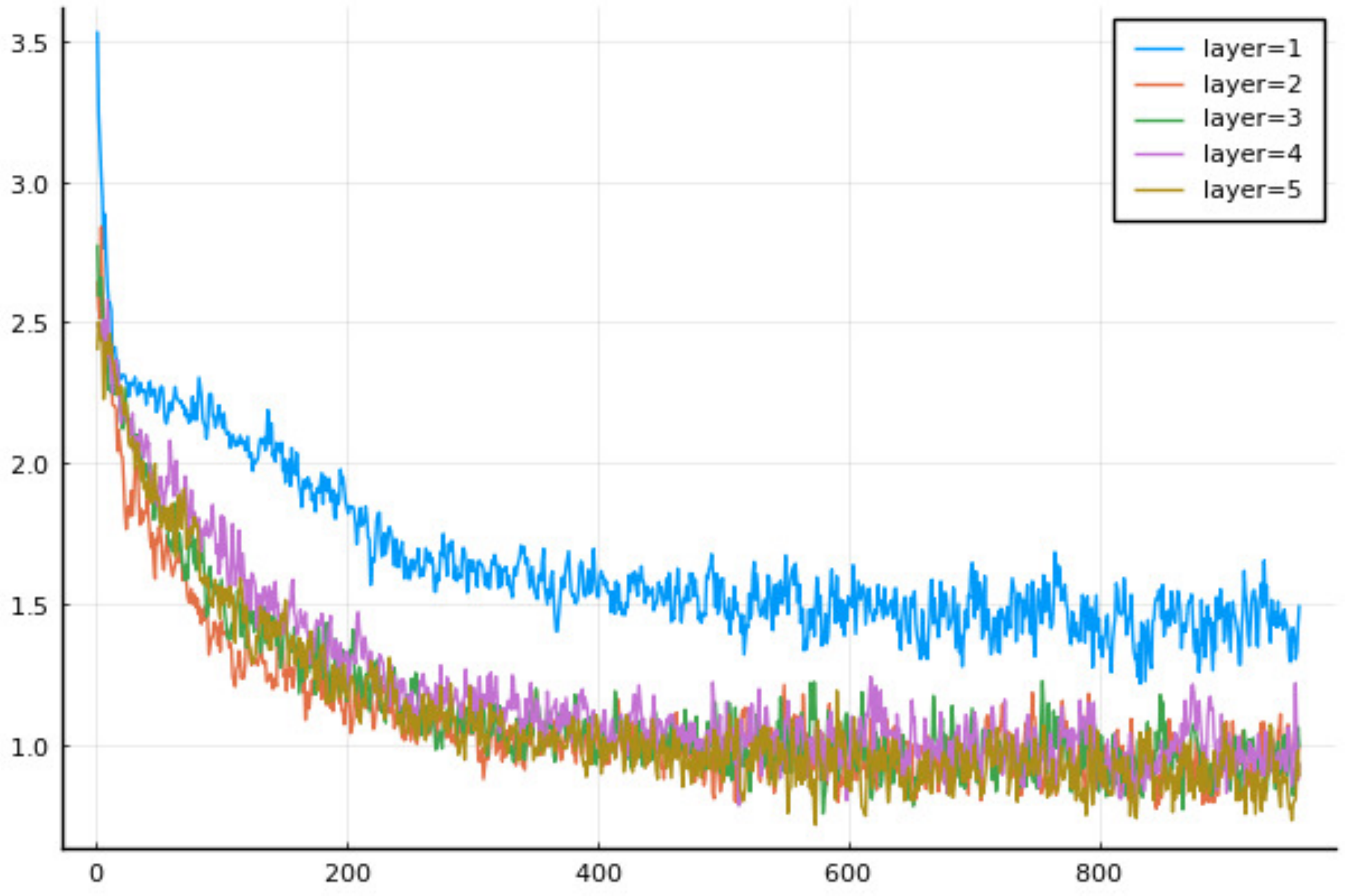} &
  \includegraphics[scale=0.27]{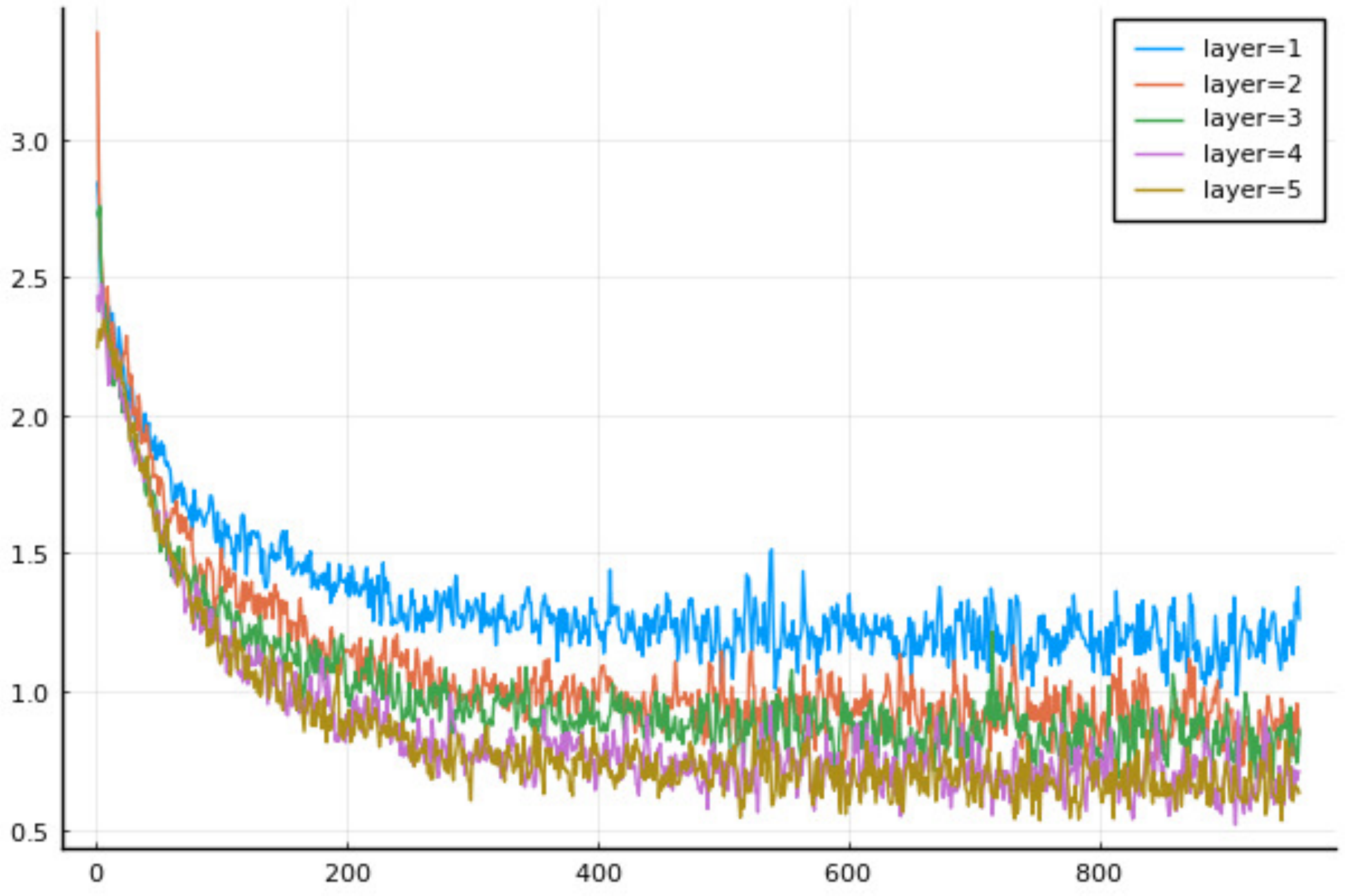} &
  \includegraphics[scale=0.27]{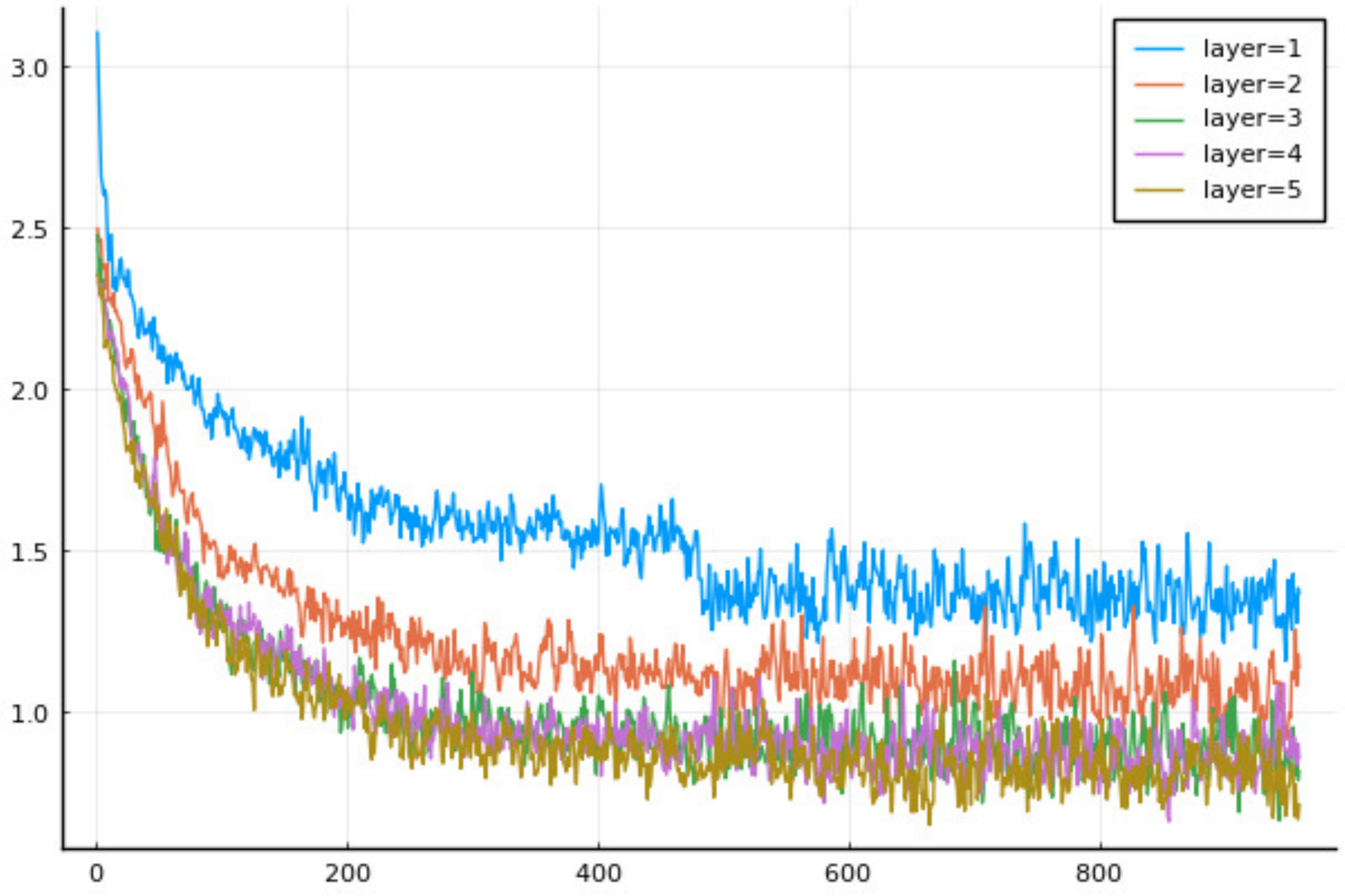}\\
  (a) costs of circuit 1 & (c) costs of circuit 2 & (e) costs of circuit 9 \\
  \\
  \includegraphics[scale=0.27]{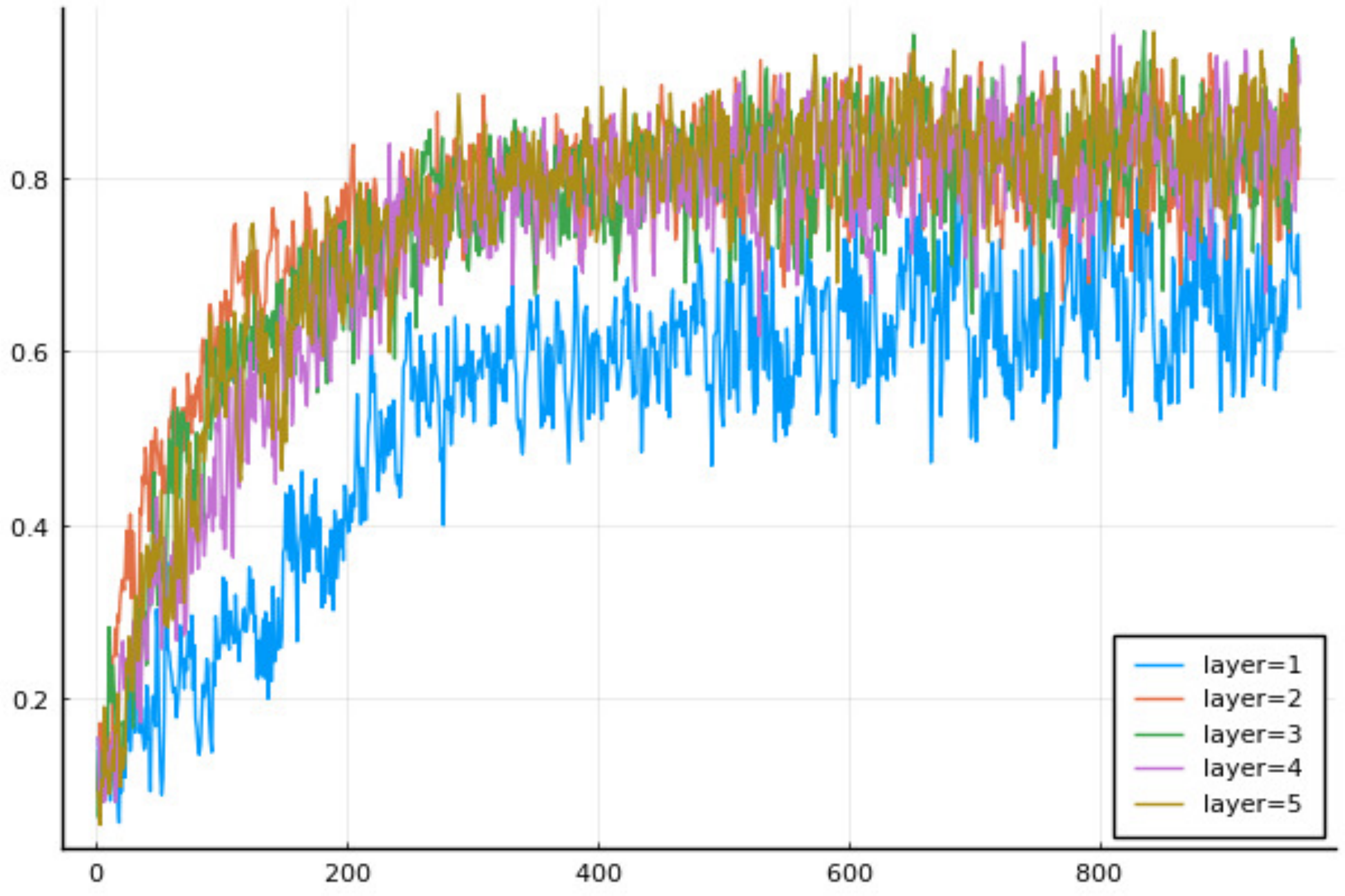} &
  \includegraphics[scale=0.27]{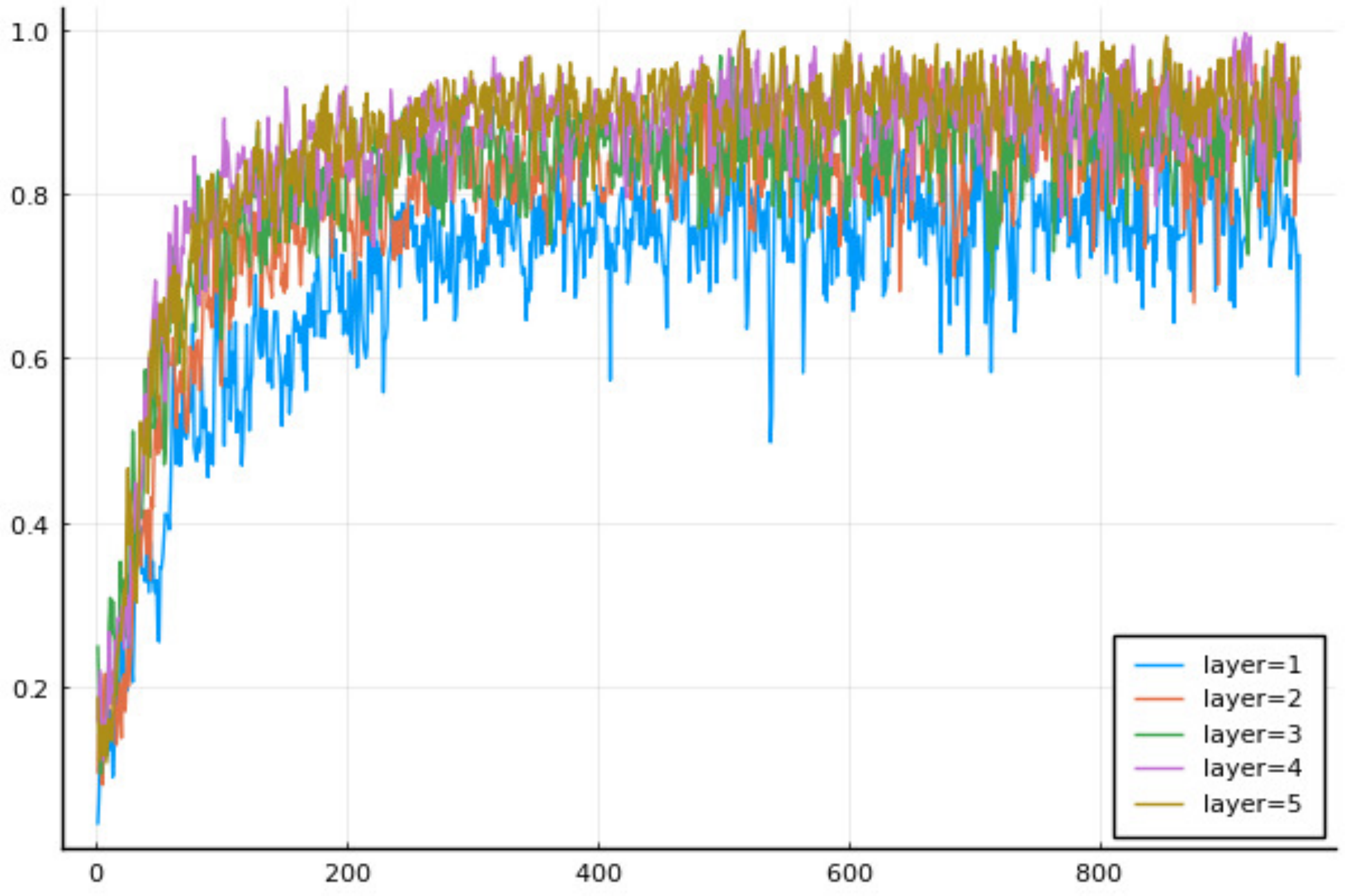} &
  \includegraphics[scale=0.27]{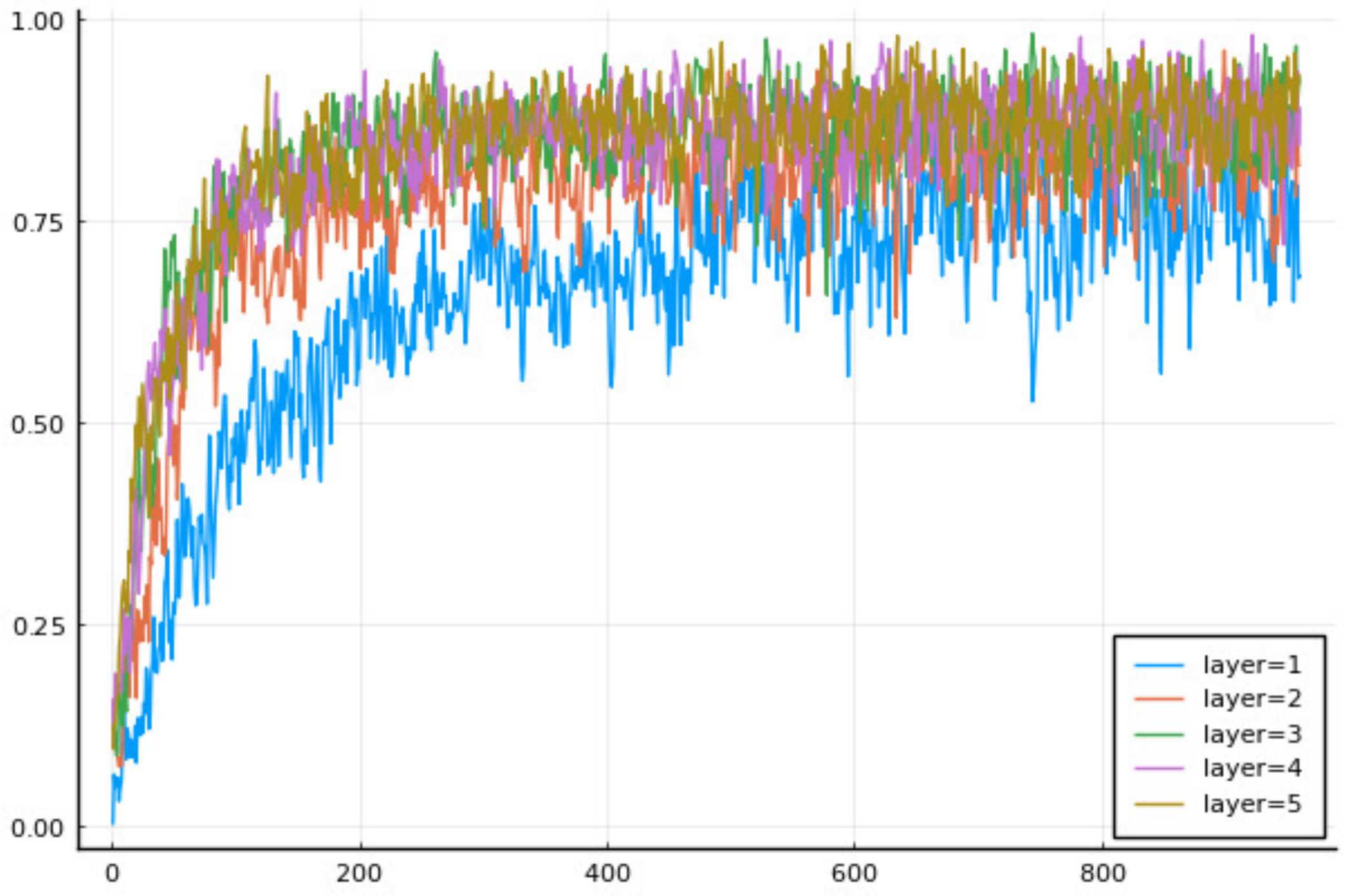}\\
  (b) accuracies of circuit 1 & (d) accuracies of circuit 2 & (f) accuracies of circuit 9 \\
  \\
  \includegraphics[scale=0.27]{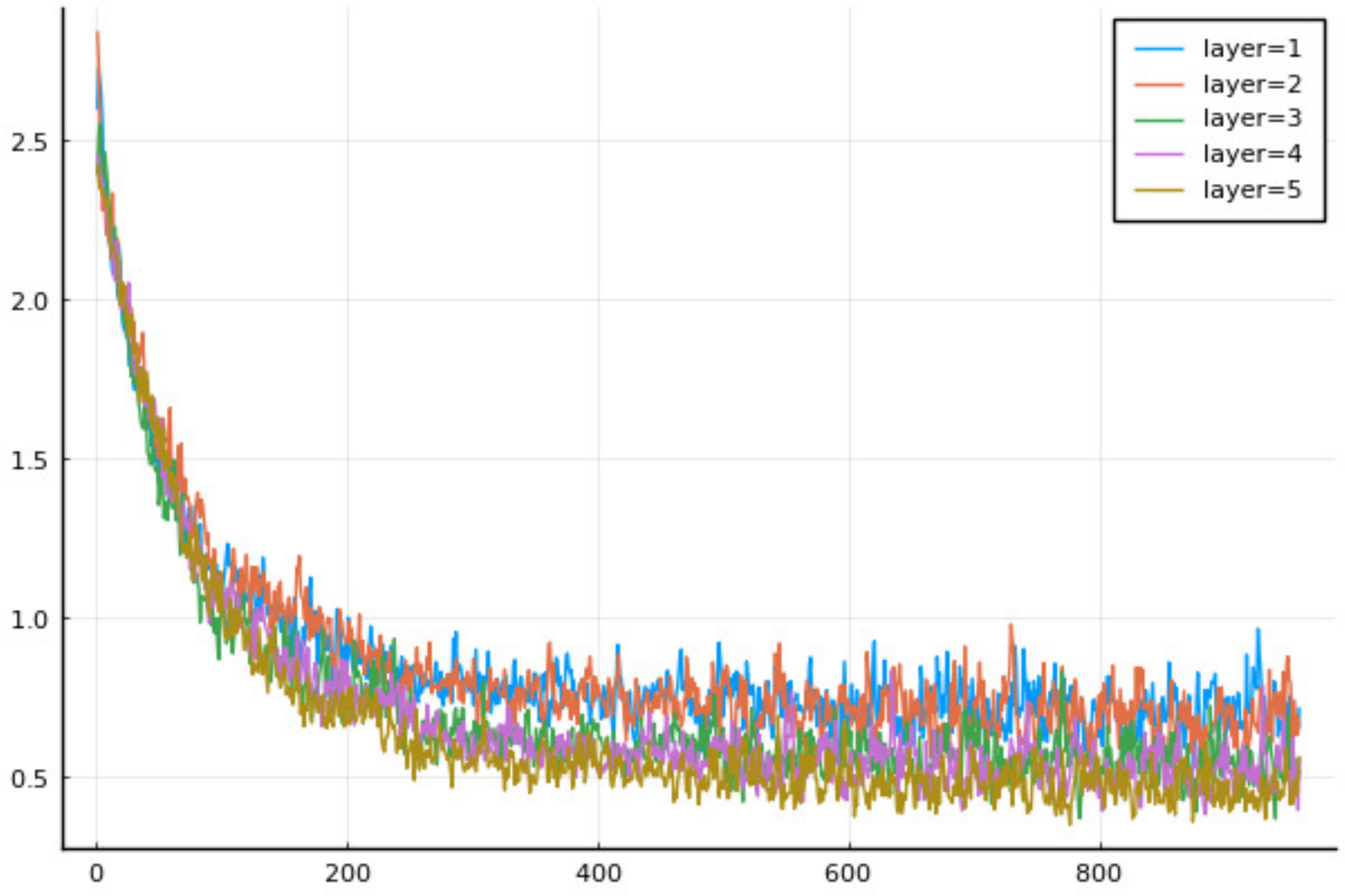} &
  \includegraphics[scale=0.27]{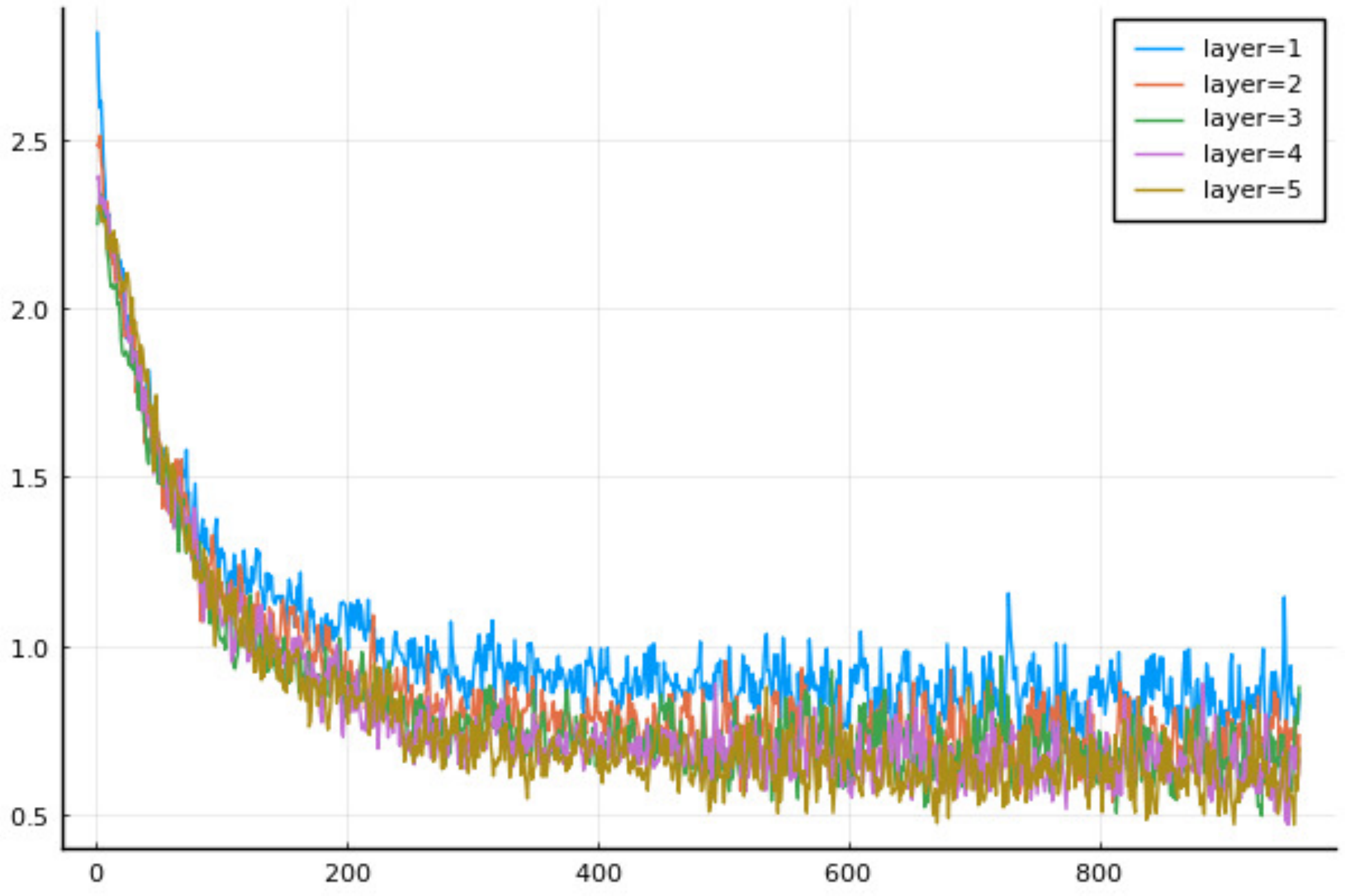} &
  \includegraphics[scale=0.27]{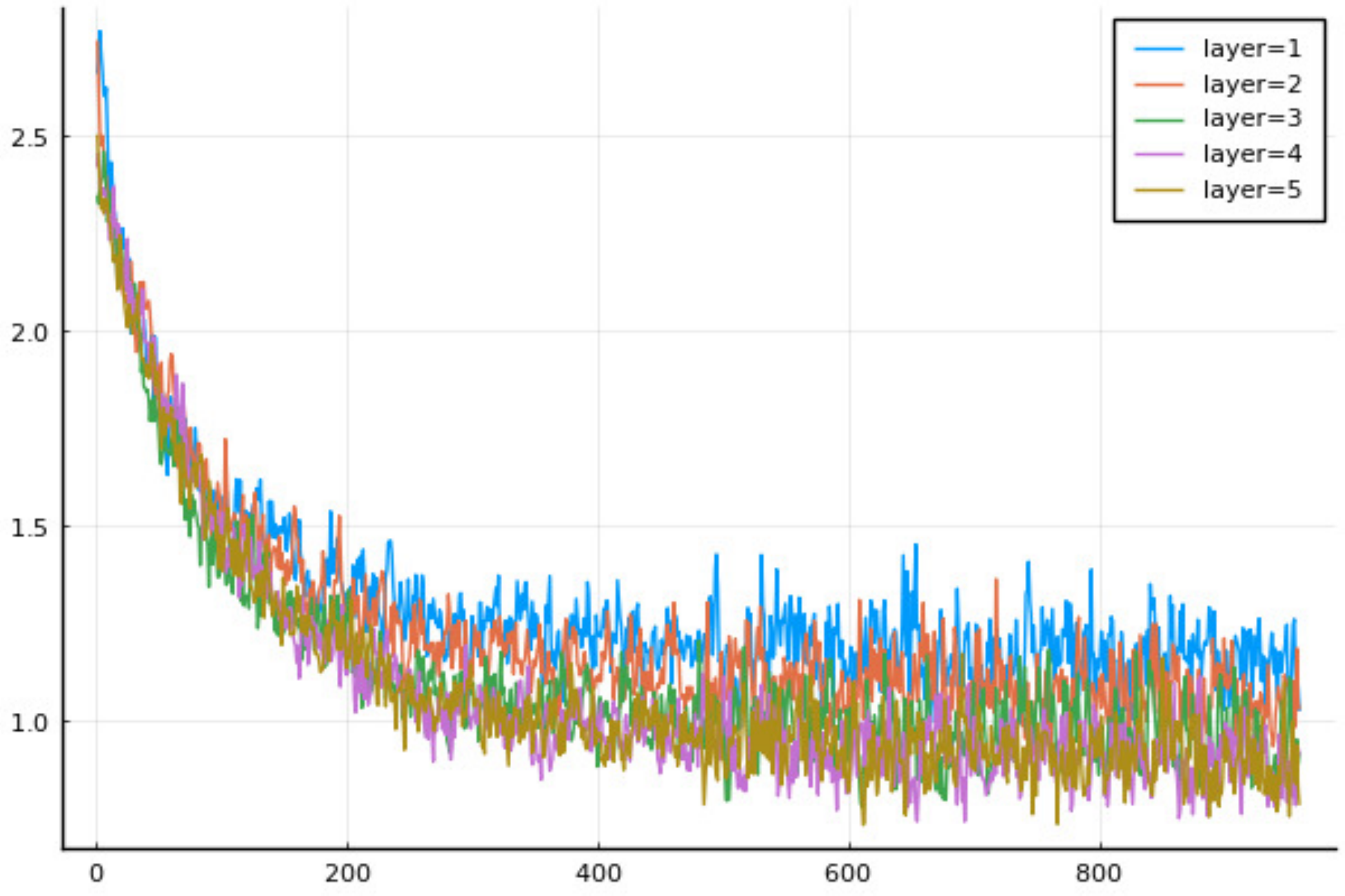}\\
  (g) costs of circuit 14 & (i) costs of circuit 15 & (k) costs of QAOA-like circuit \\
  \\
  \includegraphics[scale=0.27]{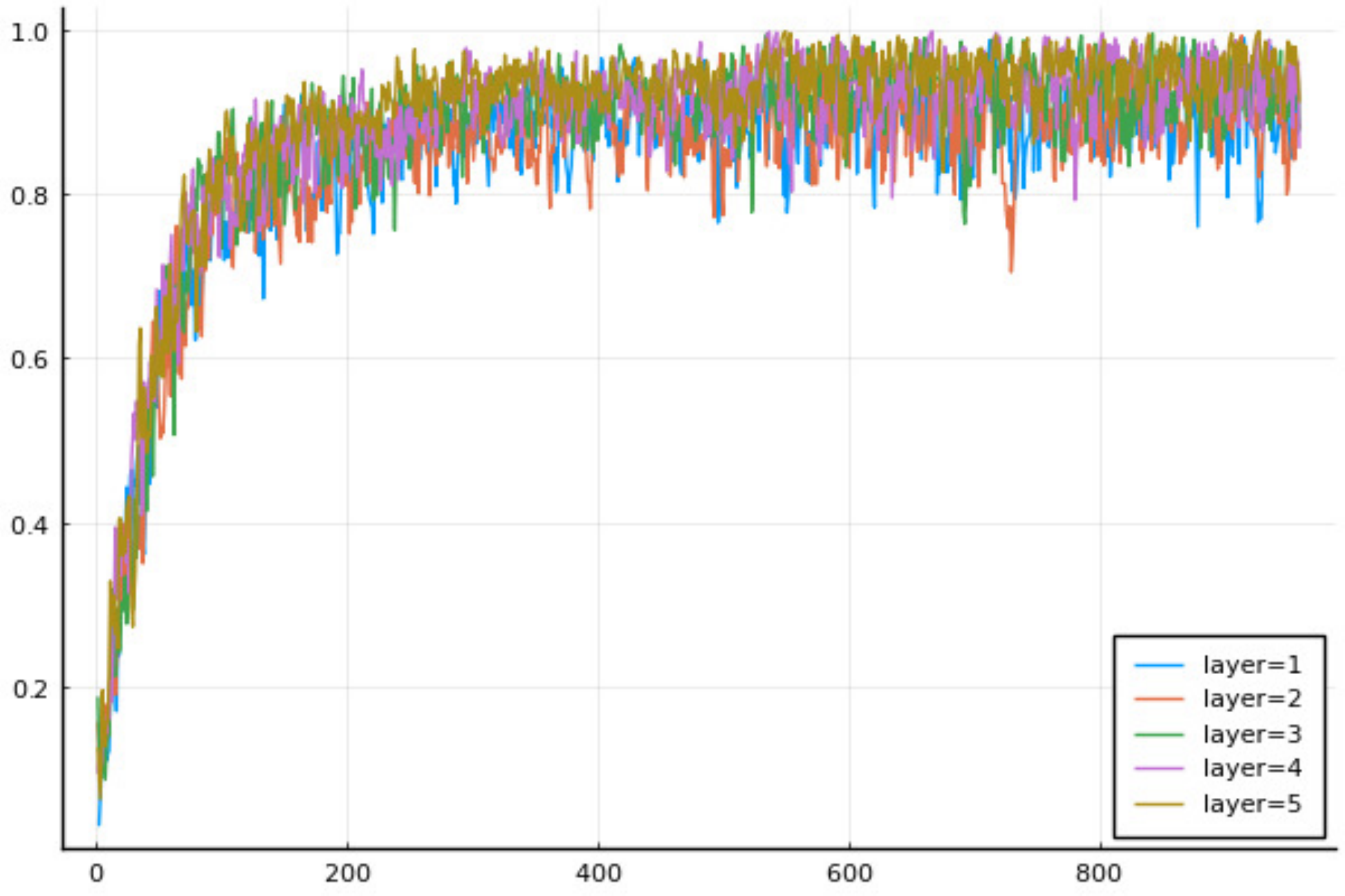} &
  \includegraphics[scale=0.27]{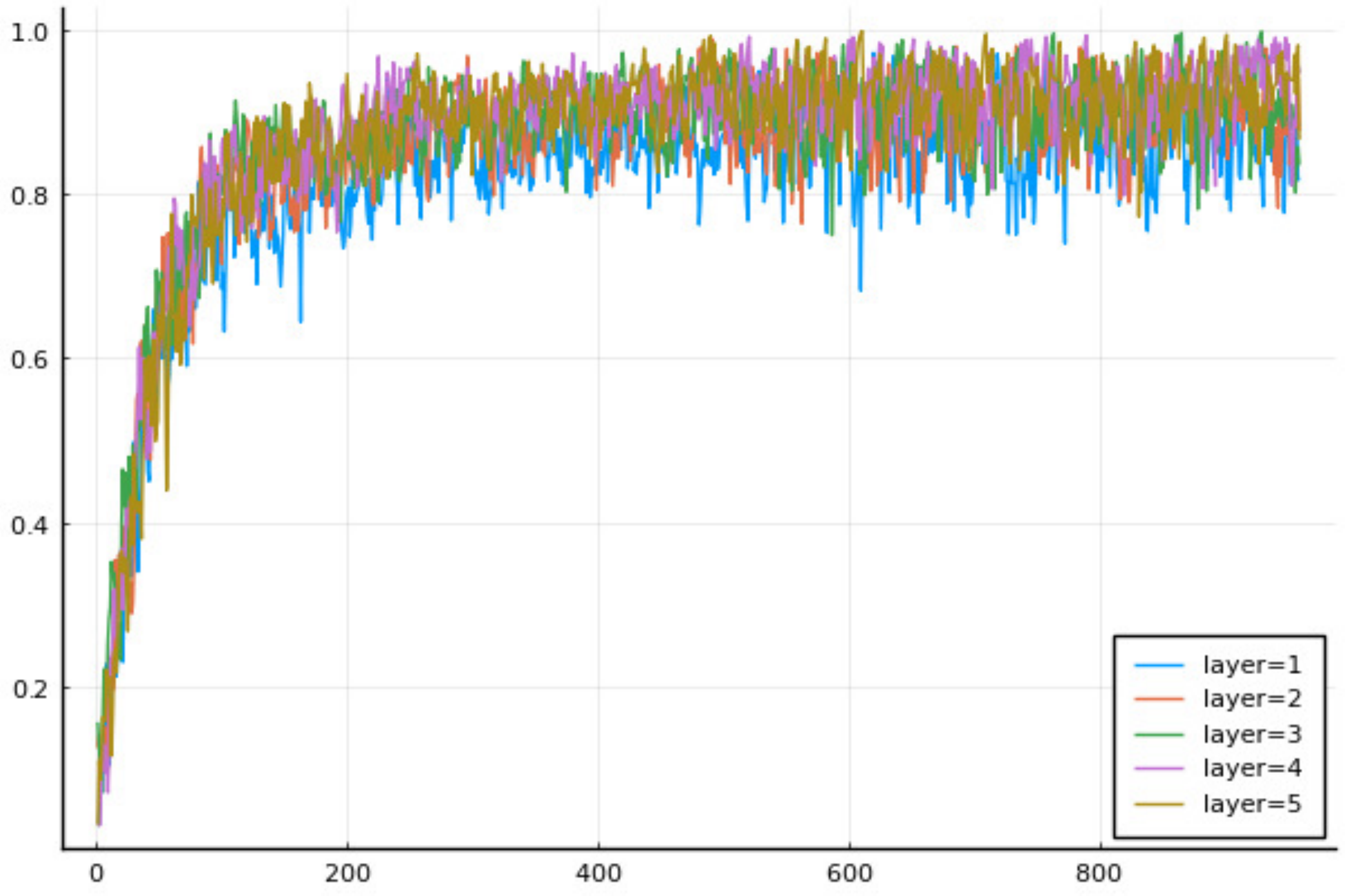} &
  \includegraphics[scale=0.27]{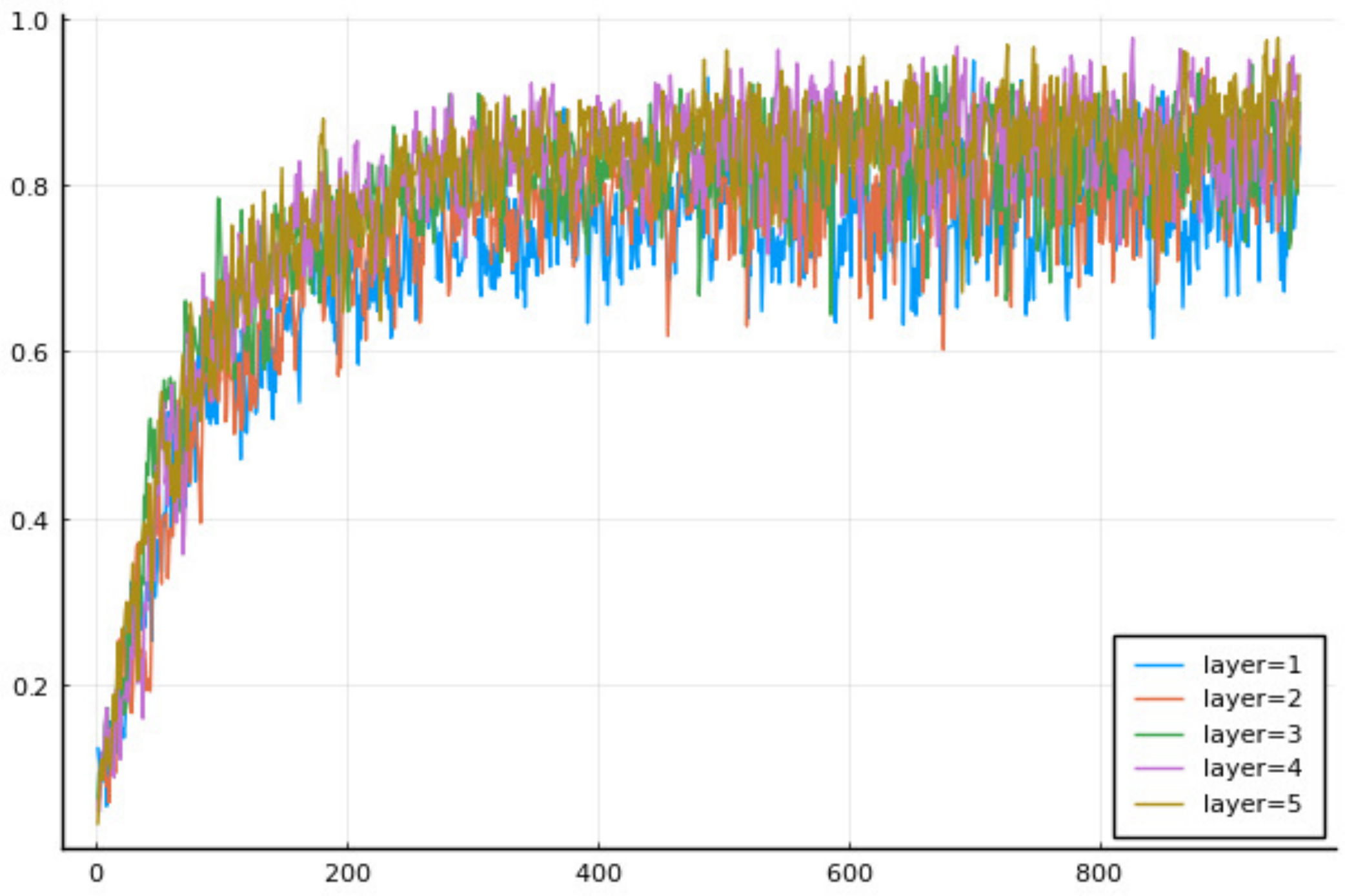}\\
  (h) accuracies of circuit 14 & (j) accuracies of circuit 15 & (l) accuracies of QAOA-like circuit \\
  \end{tabular}
  \end{center}
  \caption{The performances (cost and accuracy) of models with different settings}
  \label{Fig:F6}
\end{figure*}

\section{Conclusions}
This paper proposed a hybrid quantum-classical network with deep architectures. The new structure consists of QFE layers 
which use PQCs to ``convolve'' the input. QFE layers generalize convolution to the high dimensional hilbert space and model 
the local patches better. It has been shown that our hybrid models can achieve competitive performances based on 
simple architectures. Besides, we compare the performances of models with different ansaetze in different depths, showing that 
the model with ansatz in high expressibility performs better. 
In practice, the network architecture and the initialization method have a significant impact on performance. However, 
there is a no effective method to avoid the barren plateaus yet. On the other 
hand, since the modern classical network are so deep and our method provides many possibilities via PQCs, we cannot perform 
exhaustive tests to find the best sequence of QFE layers and the best combination of circuit ansaetze. Due to the large 
number of choices of PQCs, we cannot perform a brute force search for all the possibilities, while this opens a new space 
for the construction of hybrid quantum-classical networks. We expect the introduction of QFE layers in more architectures and extensive 
hyperparameter searches can improve the performance.

{
Besides, it is noted that quantum neural tangent kernel (QNTK) theory\cite{nakaji2021quantum, shirai2021quantum, liu2021representation} 
is developed recently, which can be applied to QNNs. It will be interesting to analyze the model with QFE layers based on 
the QNTK theory. This open question is left for future research.
}

\nocite{*}
\begin{acknowledgments}
 This work was supported by the National Natural Science Foundation of China under Grant 61873317.
\end{acknowledgments}


\begin{thebibliography}{10}%

\bibitem{arute2019quantum}
F.~Arute, K.~Arya, R.~Babbush, D.~Bacon, J.~C.~Bardin, R.~Barends, R.~Biswas, S.~Boixo, F.~G.~S.~L.~Brandao, D.~A.~Buell 
et~al.,
\newblock Quantum supremacy using a programmable superconducting processor,
\newblock Nature (London) \textbf{574}, 505 (2019).

\bibitem{preskill2018quantum}
J.~Preskill,
\newblock Quantum computing in the NISQ era and beyond,
\newblock Quantum \textbf{2}, 79 (2018).

\bibitem{biamonte2017quantum}
J.~Biamonte, P.~Wittek, N.~Pancotti, P.~Rebentrost, N.~Wiebe, and S.~Lloyd,
\newblock Quantum machine learning,
\newblock Nature (London) \textbf{549}, 195 (2017).

\bibitem{rebentrost2014quantum}
P.~Rebentrost, M.~Mohseni, and S.~Lloyd,
\newblock Quantum Support Vector Machine for Big Data Classification,
\newblock Phys. Rev. Lett. \textbf{113}, 130503 (2014).

\bibitem{lloyd2014quantum}
S.~Lloyd, M.~Mohseni, and P.~Rebentrost,
\newblock Quantum principal component analysis,
\newblock Nat. Phys. \textbf{10}, 631 (2014).

\bibitem{amin2018quantum}
M.~H.~Amin, E.~Andriyash, J.~Rolfe, B.~Kulchytskyy, and R.~Melko,
\newblock Quantum Boltzmann Machine,
\newblock Phys. Rev. X \textbf{8}, 021050 (2018).

\bibitem{mari2020transfer}
A.~Mari, T.~R.~Bromley, J.~Izaac, M.~Schuld, and N.~Killoran,
\newblock Transfer learning in hybrid classical-quantum neural networks,
\newblock Quantum \textbf{4}, 340 (2020).

\bibitem{havlivcek2019supervised}
V.~Havl{\'\i}{\v{c}}ek, A.~D.~C{\'o}rcoles, K.~Temme, A.~W.~Harrow, A.~Kandala, J.~M.~Chow, and J.~M.~Gambetta,
\newblock Supervised learning with quantum-enhanced feature spaces,
\newblock Nature (London) \textbf{567}, 209 (2019).

\bibitem{schuld2019quantum}
M.~Schuld and N.~Killoran,
\newblock Quantum Machine Learning in Feature Hilbert Spaces,
\newblock Phys. Rev. Lett. \textbf{122}, 040504 (2019).

\bibitem{benedetti2019parameterized}
M.~Benedetti, E.~Lloyd, S.~Sack, and M.~Fiorentini,
\newblock Parameterized quantum circuits as machine learning models,
\newblock Quantum Sci. Technol. \textbf{4}, 043001 (2019).

\bibitem{shor1999polynomial}
P.~W.~Shor,
\newblock Polynomial-time algorithms for prime factorization and discrete logarithms on a quantum computer,
\newblock SIAM Rev. \textbf{41}, 303 (1999).

\bibitem{grover1997quantum}
L.~K.~Grover,
\newblock Quantum Mechanics Helps in Searching for a Needle in a Haystack,
\newblock Phys. Rev. Lett. \textbf{79}, 325 (1997).

\bibitem{harrow2009quantum}
A.~W.~Harrow, A.~Hassidim, and S.~Lloyd,
\newblock Quantum Algorithm for Linear Systems of Equations,
\newblock Phys. Rev. Lett. \textbf{103}, 150502 (2009).

\bibitem{peruzzo2014variational}
A.~Peruzzo, J.~McClean, P.~Shadbolt, M.-H.~Yung, X.-Q.~Zhou, P.~J.~Love, A.~Aspuru-Guzik, and J.~L.~O’Brien,
\newblock A variational eigenvalue solver on a photonic quantum processor,
\newblock Nat. Commun. \textbf{5}, 4213 (2014).

\bibitem{wecker2015progress}
D.~Wecker, M.~B.~Hastings, and M.~Troyer,
\newblock Progress towards practical quantum variational algorithms,
\newblock Phys. Rev. A \textbf{92}, 042303 (2015).

\bibitem{mcclean2016theory}
J.~R.~McClean, J.~Romero, R.~Babbush, and A.~Aspuru-Guzik,
\newblock The theory of variational hybrid quantum-classical algorithms,
\newblock New J. Phys. \textbf{18}, 023023 (2016).

\bibitem{farhi2014quantum}
E.~Farhi, J.~Goldstone, and S.~Gutmann,
\newblock A quantum approximate optimization algorithm,
\newblock arXiv:1411.4028.

\bibitem{hadfield2019quantum}
S.~Hadfield, Z.~Wang, B.~O’Gorman, E.~G.~Rieffel, D.~Venturelli, and R.~Biswas,
\newblock From the quantum approximate optimization algorithm to a quantum alternating operator ansatz,
\newblock Algorithms \textbf{12}, 34 (2019).

\bibitem{schuld2014quest}
M.~Schuld, I.~Sinayskiy, and F.~Petruccione,
\newblock The quest for a quantum neural network,
\newblock Quantum Inf. Process. \textbf{13}, 2567 (2014).

\bibitem{farhi2018classification}
E.~Farhi and H.~Neven,
\newblock Classification with quantum neural networks on near term processors,
\newblock arXiv:1802.06002.

\bibitem{lloyd2020quantum}
S.~Lloyd, M.~Schuld, A.~Ijaz, J.~Izaac, and N.~Killoran,
\newblock Quantum embeddings for machine learning,
\newblock arXiv:2001.03622.

\bibitem{cong2019quantum}
I.~Cong, S.~Choi, and M.~D.~Lukin,
\newblock Quantum convolutional neural networks,
\newblock Nat. Phys. \textbf{15}, 1273 (2019).

\bibitem{henderson2020quanvolutional}
M.~Henderson, S.~Shakya, S.~Pradhan, and T.~Cook,
\newblock Quanvolutional neural networks: powering image recognition with quantum circuits,
\newblock Quantum Machine Intelligence \textbf{2}, 1 (2020).

\bibitem{Kerenidis2020Quantum}
I.~Kerenidis, J.~Landman, and A.~Prakash,
\newblock Quantum algorithms for deep convolutional neural networks,
\newblock In {\em Proceedings of the International Conference on Learning Representations} (ICLR, Addis Ababa, Ethiopia, 2020).

\bibitem{li2020quantum}
Y.-C.~Li, R.-G.~Zhou, R.-Q.~Xu, J.~Luo, and W.-W.~Hu,
\newblock A quantum deep convolutional neural network for image recognition,
\newblock Quantum Sci. Technol. \textbf{5}, 044003 (2020).

\bibitem{zheng2021speeding}
H.~Zheng, Z.~Li, J.~Liu, S.~Strelchuk, and R.~Kondor,
\newblock Speeding up learning quantum states through group equivariant convolutional quantum ans $\{$$\backslash$" a$\}$ tze,
\newblock arXiv:2112.07611.

\bibitem{lin2013network}
M.~Lin, Q.~Chen, and S.~Yan,
\newblock Network in network,
\newblock In {\em Proceedings of the International Conference on Learning Representations} (ICLR, Banff, AB, Canada, 2014).

\bibitem{schuld2021effect}
M.~Schuld, R.~Sweke, and J.~J.~Meyer,
\newblock Effect of data encoding on the expressive power of variational quantum machine learning models,
\newblock Phys. Rev. A \textbf{103}, 032430 (2021).

\bibitem{goto2021universal}
T.~Goto, Q.~H.~Tran, and K.~Nakajima,
\newblock Universal Approximation Property of Quantum Machine Learning Models in Quantum-Enhanced Feature Spaces,
\newblock Phys. Rev. Lett. \textbf{127}, 090506 (2021).

\bibitem{liu2021hybrid}
J.~Liu, K.~H.~Lim, K.~L.~Wood, W.~Huang, C.~Guo, and H.-L.~Huang,
\newblock Hybrid quantum-classical convolutional neural networks,
\newblock Sci. China Phys. Mech. Astron. \textbf{64}, 290311 (2021).

\bibitem{lecun1989backpropagation}
Y.~LeCun, B.~Boser, J.~S.~Denker, D.~Henderson, R.~E.~Howard, W.~Hubbard, and L.~D.~Jackel,
\newblock Backpropagation applied to handwritten zip code recognition,
\newblock Neural Comput. \textbf{1}, 541 (1989).

\bibitem{lecun1998gradient}
Y.~LeCun, L.~Bottou, Y.~Bengio, and P.~Haffner,
\newblock Gradient-based learning applied to document recognition,
\newblock Proc. IEEE \textbf{86}, 2278 (1998).

\bibitem{krizhevsky2012imagenet}
A.~Krizhevsky, I.~Sutskever, and G.~E.~Hinton,
\newblock Imagenet classification with deep convolutional neural networks,
\newblock In {\em Advances in Neural Information Processing Systems} (NeurIPS, Lake Tahoe, NV, USA, 2012), pp. 1097-1105.

\bibitem{redmon2016you}
J.~Redmon, S.~Divvala, R.~Girshick, and A.~Farhadi,
\newblock You only look once: Unified, real-time object detection,
\newblock In {\em Proceedings of the IEEE Conference on Computer Vision and Pattern Recognition} (CVPR, Las Vegas, NV, USA, 2016), pp. 779-788.

\bibitem{long2015fully}
J.~Long, E.~Shelhamer, and T.~Darrell,
\newblock Fully convolutional networks for semantic segmentation,
\newblock In {\em Proceedings of the IEEE Conference on Computer Vision and Pattern Recognition} (CVPR, Boston, MA, USA, 2015), pp. 3431-3440.

\bibitem{zhang2016faster}
L.~Zhang, L.~Lin, X.~Liang, and K.~He,
\newblock Is faster R-CNN doing well for pedestrian detection?,
\newblock In {\em European Conference on Computer Vision} (ECCV, Amsterdam, The Netherlands, 2016), pp. 443-457.

\bibitem{vaswani2017attention}
A.~Vaswani, N.~Shazeer, N.~Parmar, J.~Uszkoreit, L.~Jones, A.~N.~Gomez, {\L}.~Kaiser, and I.~Polosukhin,
\newblock Attention is all you need,
\newblock In {\em Advances in Neural Information Processing Systems} (NeurIPS, Long Beach, CA, USA, 2017), pp. 5998-6008.

\bibitem{szegedy2015going}
C.~Szegedy, W.~Liu, Y.~Jia, P.~Sermanet, S.~Reed, D.~Anguelov, D.~Erhan, V.~Vanhoucke, and A.~Rabinovich,
\newblock Going deeper with convolutions,
\newblock In {\em Proceedings of the IEEE Conference on Computer Vision and Pattern Recognition} (CVPR, Boston, MA, USA, 2015), pp. 1-9.

\bibitem{simonyan2014very}
K.~Simonyan and A.~Zisserman,
\newblock Very deep convolutional networks for large-scale image recognition,
\newblock In {\em Proceedings of the International Conference on Learning Representations} (ICLR, San Diego, CA, USA, 2015).

\bibitem{he2016deep}
K.~He, X.~Zhang, S.~Ren, and J.~Sun,
\newblock Deep residual learning for image recognition,
\newblock In {\em Proceedings of the IEEE Conference on Computer Vision and Pattern Recognition} (CVPR, Las Vegas, NV, USA, 2016), pp. 770-778.

\bibitem{huang2017densely}
G.~Huang, Z.~Liu, L.~V.~D.~Maaten, and K.~Q.~Weinberger,
\newblock Densely connected convolutional networks,
\newblock In {\em Proceedings of the IEEE Conference on Computer Vision and Pattern Recognition} (CVPR, Honolulu, HI, USA, 2017), pp. 2261-2269.

\bibitem{dallaire2018quantum}
P.-L.~Dallaire-Demers and N.~Killoran,
\newblock Quantum generative adversarial networks,
\newblock Phys. Rev. A \textbf{98}, 012324 (2018).

\bibitem{romero2017quantum}
J.~Romero, J.~P.~Olson, and A.~Aspuru-Guzik,
\newblock Quantum autoencoders for efficient compression of quantum data,
\newblock Quantum Sci. Technol. \textbf{2}, 045001 (2017).

\bibitem{sharma2020noise}
K.~Sharma, S.~Khatri, M.~Cerezo, and P.~J.~Coles,
\newblock Noise resilience of variational quantum compiling,
\newblock New J. Phys. \textbf{22}, 043006 (2020).

\bibitem{mitarai2018quantum}
K.~Mitarai, M.~Negoro, M.~Kitagawa, and K.~Fujii,
\newblock Quantum circuit learning,
\newblock Phys. Rev. A \textbf{98}, 032309 (2018).

\bibitem{lloyd2018quantum}
S.~Lloyd,
\newblock Quantum approximate optimization is computationally universal,
\newblock arXiv:1812.11075.

\bibitem{morales2020universality}
M.~E.~S.~Morales, J.~D.~Biamonte, and Z.~Zimbor{\'a}s,
\newblock On the universality of the quantum approximate optimization algorithm,
\newblock Quantum Inf. Process. \textbf{19}, 291 (2020).

\bibitem{luo2020yao}
X.-Z.~Luo, J.-G.~Liu, P.~Zhang, and L.~Wang,
\newblock Yao. jl: Extensible, efficient framework for quantum algorithm design,
\newblock Quantum \textbf{4}, 341 (2020).

\bibitem{sim2019expressibility}
S.~Sim, P.~D.~Johnson, and A.~Aspuru-Guzik,
\newblock Expressibility and entangling capability of parameterized quantum circuits for hybrid quantum-classical algorithms,
\newblock Adv. Quantum Technol. \textbf{2}, 1900070 (2019).

\bibitem{mcclean2018barren}
J.~R.~McClean, S.~Boixo, V.~N.~Smelyanskiy, R.~Babbush, and H.~Neven,
\newblock Barren plateaus in quantum neural network training landscapes,
\newblock Nat. Commun. \textbf{9}, 4812 (2018).

\bibitem{nakaji2021quantum}
K.~Nakaji, H.~Tezuka, and N.~Yamamoto,
\newblock Quantum-enhanced neural networks in the neural tangent kernel framework,
\newblock arXiv:2109.03786.

\bibitem{shirai2021quantum}
N.~Shirai, K.~Kubo, K.~Mitarai, and K.~Fujii,
\newblock Quantum tangent kernel,
\newblock arXiv:2111.02951.

\bibitem{liu2021representation}
J.~Liu, F.~Tacchino, J.~R.~Glick, L.~Jiang, and A.~Mezzacapo,
\newblock Representation learning via quantum neural tangent kernels,
\newblock arXiv:2111.04225.

\end{thebibliography}

%

\end{document}